\documentclass[10pt]{olplainarticle}

\usepackage{adjustbox}
\usepackage{threeparttable}
\usepackage{booktabs}   
\usepackage[section]{placeins}
\usepackage{float}
\usepackage{amssymb}
\usepackage[table]{xcolor}
\usepackage{caption}
\usepackage{threeparttable}
\usepackage{adjustbox}
\usepackage{bm}

\title{Regularized joint reconstruction and slab combination for accelerated three-dimensional multi-slab diffusion-weighted imaging using multi-scale energy models}

\author[1]{Reza Ghorbani}
\author[1]{Jyothi Rikhab Chand}
\author[2]{Chu-Yu Lee}
\author[1]{Mathews Jacob}
\author[3]{Merry Mani}

\affil[1]{Department of Electrical and Computer Engineering, University of Virginia, Virginia, USA}
\affil[2]{Department of Radiology, University of Iowa, Iowa, USA}
\affil[3]{Department of Radiology and Medical Imaging, University of Virginia, Virginia, USA}


\begin{abstract}
This work presents Energy-based Profile Encoding, EPEN, a joint reconstruction framework for high-resolution diffusion-weighted MRI from undersampled 3D multi-slab k-space acquisitions, designed to suppress slab-boundary artifacts while preserving fine anatomical detail. EPEN formulates the multi-slab acquisition process using a bilinear forward model in which both the diffusion-weighted image volume and slab excitation profiles are treated as unknown variables. Reconstruction is posed as a maximum a posteriori optimization problem with three components: a Gaussian data-fidelity term enforcing consistency with the acquired k-space measurements, a CNN-based deep energy prior that represents the negative log distribution of clean diffusion-weighted images, and a quadratic regularization term that constrains the estimated slab profiles toward an initial profile estimate. The gradient of the learned energy prior guides accelerated reconstruction toward an artifact-free image distribution. The resulting nonconvex objective is solved using alternating minimization, with image-volume updates performed through a majorize-minimize scheme using conjugate-gradient optimization and slab-profile updates estimated by regularized least squares. Across multiple acceleration factors and slab configurations, EPEN substantially reduced slab-boundary artifacts compared with conventional slab-boundary correction methods, while improving structural consistency and preserving diffusion-weighted contrast. These results demonstrate that EPEN enables robust joint 3D multi-slab diffusion MRI reconstruction and slab-profile correction within a unified optimization framework supported by deep energy-based image priors.
\end{abstract}

\begin{document}

\flushbottom
\maketitle
\thispagestyle{empty}

\section{Introduction}

Diffusion-weighted MRI (dMRI) is known for its exceptional sensitivity to encode water displacements at the microscopic scales, making it a powerful tool for detecting early cellular changes in the brain tissue. When coupled with high spatial resolution, dMRI enables precise localization of microstructural alterations, facilitating the study of small and complex neuroanatomical structures and advancing our understanding of brain development, pathology, and connectivity.  

 3D multi-slab diffusion-weighted echo-planar imaging (DW-EPI) offers an SNR-efficient approach for high-resolution diffusion imaging\cite{Engstrom2013, Frost2014a, Bruce2017, Dai2021}. In this technique, the full imaging volume is divided into multiple slabs, and a multi-shot readout is employed to sample the k$_z$ dimension of each slab to resolve slices. With the slabs excited and encoded successively (or in a simultaneous multi-slab fashion), the repetition time (TR) remains within the optimal range of 1–2 seconds for spin-echo based acquisitions at 3T field strength. Notably, this TR of 1-2 sec per k$_z$ encode can lead to long total volume acquisition time (VAT) for high resolution applications\cite{Moeller2020}. This becomes particularly problematic for studies requiring high angular or q-space resolution, such as advanced microstructural imaging or fiber tractography, where scan time scales linearly with the number of volumes and the k$_z$ shots used per volume. To reduce long VAT, several accelerated 3D k-space acquisition methods have been proposed to use fewer $k_z$ shots per volume acquisition\cite{Lee2024, Li2023, Li2024}. Such acceleration methods are crucial for enabling advanced high angular/q-space resolution applications. Recently, thicker slabs with more slices have been shown to achieve higher acceleration factors, especially for high b-value applications, capitalizing on incoherent sampling patterns using 2D-CAIPI-based approaches coupled with an extended field-of-view approach\cite{Lee2026}.
 
 The high imaging fidelity and higher SNR efficiency offered by 3D multi-slab acquisition can be compromised by slab boundary artifacts, an inevitable consequence of the 3D multi-slab excitation \cite{Engstrom2013}. These artifacts arise primarily from the truncation of the theoretically infinite radiofrequency (RF) pulse required to achieve a perfectly rectangular excitation profile. In practice, the truncated RF pulse results in a non-ideal excitation profile characterized by magnitude variations within the main lobe, non-zero side lobes, and extended transition bands. These imperfections lead to signal intensity fluctuations within slabs, and excitation of regions beyond the intended slab boundaries. The slab crosstalk, arising from the unintended excitation of adjacent slabs, and slab aliasing, arising from excitation extending beyond the intended slab thickness, lead to slab boundary artifacts. The severity of these artifacts is further exacerbated in VAT-accelerated acquisitions, where undersampling of the 3D k-space amplifies the slab boundary artifacts.

 To mitigate slab boundary artifacts while combining the slabs, oversampling along the slice direction is typically performed. However, oversampling can result in an increased number of $k_z$ encodings and hence increased scan-time. To address this inefficiency, a calibration-based slab-combination method known as PEN \cite{van2015slab} was proposed. In PEN, a calibration scan with oversampling in the slice direction is used to estimate the extended RF excitation profile. This profile is applied to non-oversampled imaging data, enabling artifact-free slab combination through a linear slice-unmixing inversion process analogous to SENSE reconstruction \cite{pruessmann1999sense}. PEN maintains slab overlap while avoiding oversampling during the main acquisition, thereby reducing scan time without compromising image quality. Building on this framework, the NPEN method treats both the slab-combined volume and the slab excitation profiles as unknowns, solving them simultaneously using a non-linear inversion algorithm \cite{wu2016reducing}. The initial guess for the NPEN method is derived from an over-sampled volume (typically a b0 volume), while the non-b0 images are not over-sampled but overlapped, similar to the PEN setting. With iterative updates to the slab profiles, NPEN additionally corrects for slab cross-talk effects, providing improved performance, especially at TR close to 2 seconds. Both PEN and NPEN effectively correct slab boundary artifacts, with the calibration pre-scan introducing only minimal scan time overhead.
More recently, CPEN \cite{zhang2022slab} addressed the high computational cost of NPEN by replacing its iterative solver with a CNN-based update within the same joint image–slab profile inversion, substantially accelerating the reconstruction and improving performance.

We note that all these methods are designed for fully sampled acquisitions along the slice direction, where the available slab profile sufficiently encodes slice-specific structures. Under $k_z$ accelerated scenarios, the encoding becomes ill-conditioned, resulting in residual slab artifacts. While PEN employs no explicit regularization, NPEN makes use of hand-crafted regularizations that suppress frequencies where slab artifacts manifest. Such regularization priors may not be sufficient to suppress slab-profile errors in accelerated along $k_z$ cases. 

Recent approaches exploring novel sampling patterns to accelerate 3D k-space open new possibilities for distributing sampling patterns and combining slabs. We address the limitations of the previous methods by framing image recovery as a regularized joint reconstruction and slab combination, enabling the use of 3D k-space undersampling patterns.

More recently, machine learning–based approaches have employed plug-and-play (PnP) denoisers modeled using neural networks (NNs). In contrast to traditional PnP methods, which implicitly define regularization through a denoiser, we adopt an explicit energy-based prior within a maximum a posteriori (MAP) framework. Specifically, the prior is modeled as a CNN-based energy function trained independently, whose gradient with respect to the input, the score, acts as a learned regularizer during reconstruction. This formulation enables direct integration of the learned prior into the optimization while maintaining a well-defined objective function. The resulting algorithm benefits from a principled optimization framework with monotonic descent behavior, avoiding the contraction constraints typically required in classical PnP methods, which often rely on spectral normalization of network blocks to ensure stability. Furthermore, unlike conventional PnP approaches that rely on single-scale denoisers, we employ a multiscale energy model, which better captures complex image statistics and improves reconstruction performance.

The proposed energy-based slab profile encoding framework, EPEN, simultaneously achieves image recovery and noise reduction in accelerated imaging. Both the noise-suppressed slab-combined volume and artifact-free slab-profile are treated as unknowns and updated iteratively. The learned regularizer is expected to improve slab-profile estimates in accelerated imaging, aiding the slab combination process. The proposed regularization is shown to provide improved reconstruction performance for various 3D k-space sampling trajectories using experiments performed on simulated and real human data.

\section{Methods}

We first formulate the regularized joint reconstruction and slab-combination:

\subsection{Forward model Formulation}
Let $\mathbf{u} \in \mathbb{C}^{N_x\cdot  N_y\cdot N_z}$ denote the unknown complex-valued 3D slab-combined image defined on a Cartesian grid of size $N_x\times N_y\times N_z$. The total field of view (FOV) in the slice direction is covered by $N_s$ slabs. Because the slab profile can be affected by tissue and spatial variations in $B_1$ \cite{van2015slab,wu2016reducing}, we model it using 3-D functions $\mathbf S = \{ \mathbf s_i \}_{i=1}^{N_s} \in \mathbb{C}^{N_x\cdot N_y\cdot N_z\cdot N_s}$, where $i$ is the slab index. 

The measured multi-coil, multi-slab k-space data can then be modeled as 
\begin{equation}
\mathbf d=\mathcal{A}(\mathbf u,\mathbf S) + \mathbf n, \  \mathcal{A(\mathbf u,\mathbf S)}= \mathbf{PFC(S \circ u)},
\end{equation}
where $ \mathbf d$ is the vector of the measured k-space data stacked from across slabs and coils, $ \mathbf F$ is the block-diagonal 3D DFT operator, $ \mathbf C$ denotes the multiplication operator applying coil sensitivities, and $\circ$ denotes point-wise multiplication of the slab profile with the 3D slab-combined volume $\mathbf u$. $ \mathbf P$ is the sampling operator that selects the acquired k-space samples, and $\mathbf n \sim \mathcal{N}(0,\eta^2 \mathbf{I})$ is measurement noise assumed to be complex white Gaussian noise with variance $\eta^2$.

In conventional PEN/NPEN acquisitions, an extended slab profile is first estimated from an oversampled calibration scan. Then, during imaging, the encoded FOV in the slice direction, $ \Delta FOV_z$, is set to match the nominal slab thickness. Because the RF excitation profile extends beyond this encoded $\Delta FOV_z$, the signal originating outside the encoded region aliases back into the slab, producing characteristic slab fold-over artifacts. The number of $k_z$ encodes, $N_{k_{z}}$, is determined by the desired slice resolution based on this $\Delta FOV_z$. This reduction in the phase‑encoded $\Delta FOV_z$ and thereby $N_{k_{z}}$ constitutes one form of acceleration in PEN/NPEN, since avoiding slab aliasing would otherwise require oversampling the slice‑encoding direction.

In this work, we adopt an alternate strategy, decoupling slab thickness from the phase-encoded FOV in the slice direction. Specifically, we set $\Delta FOV_z > \text{desired slab thickness}$ while maintaining the same nominal slice resolution. Rather than increasing the number of acquired $k_z$ encodes, we keep $N_{k_{z}}$ fixed and realize the extended FOV by skipping a subset of $k_z$ lines according to a prescribed sampling pattern. As illustrated in Figure 1, this strategy converts the classical fold-over slab aliasing into a more distributed pattern that is easier to resolve using joint reconstruction with appropriate priors, without requiring an additional oversampled calibration scan. Further, this framework allows the introduction of arbitrary undersampling patterns for accelerated 3D multi-slab acquisitions and the use of thicker slabs, unlike the thin-slab constraints in classical settings.

\begin{figure}
  \includegraphics[width=1\linewidth,height=0.5\linewidth]{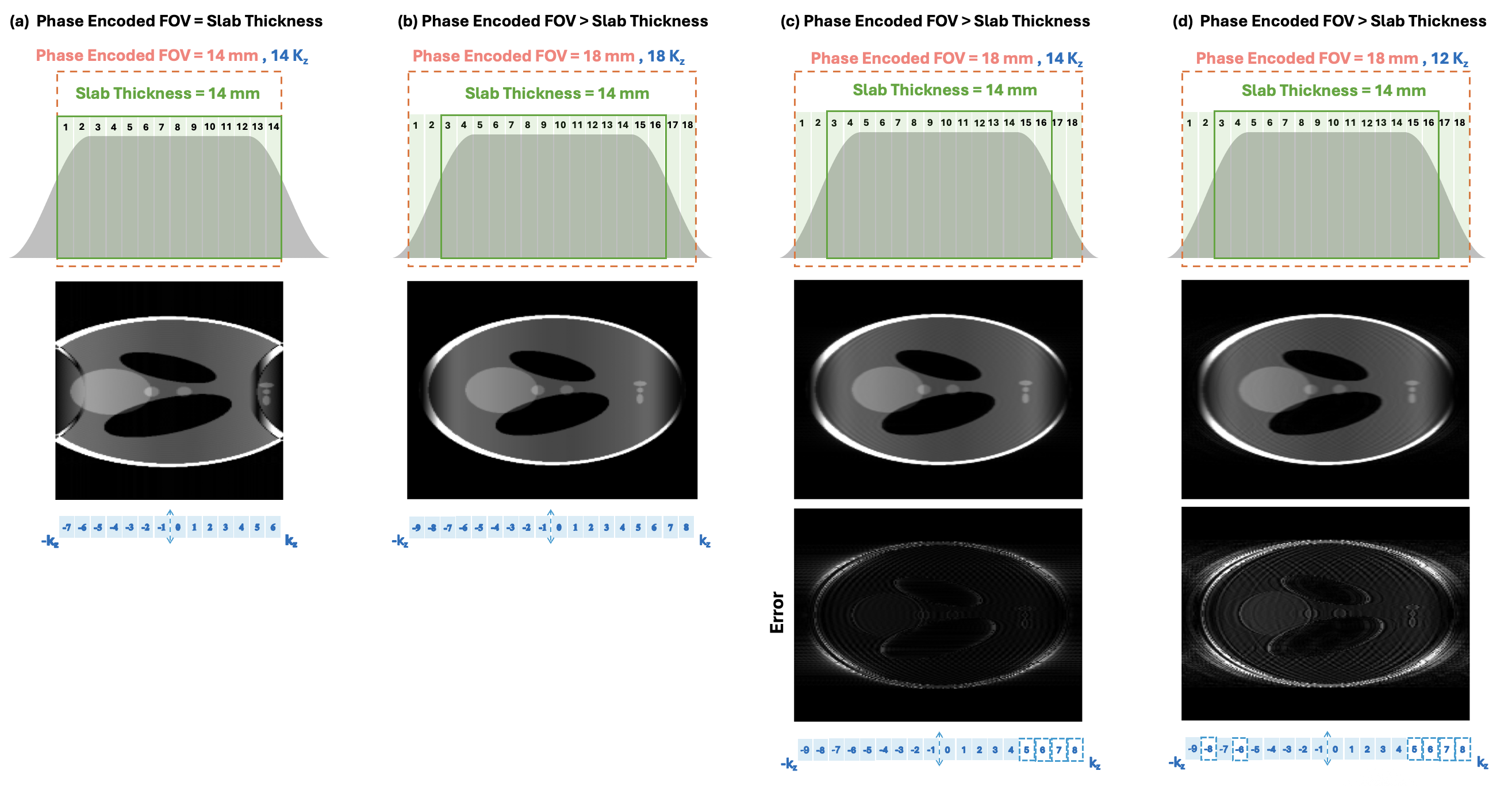}
       \caption{Schematic illustration of the modified phase-encoding. The example shown uses a prescribed slab thickness of 14mm and a desired slice thickness of 1mm along the z-direction. (a) In the classical setting, $\Delta FOV_z$ is set equal to the slab thickness, and the number of phase encodes is determined by the desired slice thickness, leading to N$_{k_z}$ of 14 for this case. Since the non-ideal RF pulse (shown in gray) extends beyond the prescribed slab thickness (shown as a green box), and $\Delta FOV_z$ does not fully cover the excited region, structured FOV fold-over artifacts will manifest in this setting as shown. In the modified extended FOV approach given in (b-d), the $\Delta FOV_z$ $>$ slab thickness, which prevents the structured FOV fold-over. To achieve the desired slice thickness, N$_{k_z}$ needs to be increased to 18 in the fully sampled setting shown in (b), which leads to increased scan-time. To reduce scan-time, accelerated approaches can be employed where $k_z$ phase-encoding is skipped, as shown in (c) with N$_{k_z}$=14, or in (d) with N$_{k_z}$=12. The k-space under-sampling patterns shown in the bottom panel, shown in blue, can be curated such that a more distributed alias pattern is formed, which is easier to solve using regularized reconstruction methods. Note that this figure is for illustration purposes only, and the depicted $k_z$ sampling parameters differ from the exact experimental sampling settings.}
\end{figure}

\subsection{Inverse problem}
MAP is a statistical framework for solving inverse problems, where the unknown is recovered by maximizing the posterior probability $p(\mathbf{u,s ~|~d})$ or, equivalently, the negative log-posterior \cite{dashti2013bayesian,dunlop2020hyperparameter}. Accordingly, in the proposed MAP framework, we recover $\mathbf u$ and $\mathbf S$ jointly by solving the following bilinear regularized optimization: 

\begin{equation}
\min_{\mathbf{u},\,\mathbf S}\;
\underbrace{\frac{1}{2 \eta^2}\bigl\lVert\mathbf{A}(\mathbf{u},\mathbf S)-\mathbf{d}\bigr\rVert_{2}^{2}}_{-log(p(d|\mathbf{u,S}))}
\;+\;\underbrace{\lambda_u\,\mathcal{R}_u(\mathbf u)\;+\lambda_S\,\mathcal{R}_S(\mathbf S)}_{-log(p(\mathbf{u,s})},
\end{equation}

\noindent where the posterior distribution, $-log(p(\mathbf{u,S ~|~d}))$, is expressed in terms of the negative log likelihood and the negative log-prior distribution of the unknowns. The data fidelity term forms the log likelihood, $-log(p(\mathbf d|\mathbf{u,S}))$, while we replace the classical negative log-prior for the image u, $\mathcal{R}_u(\mathbf u)$, with a learned energy function. When trained over a prior distribution of clean artifact-free images, this energy serves as an effective prior by favoring reconstructions that lie close to this underlying distribution. The prior distribution is modeled by a differentiable CNN which can be directly plugged into the iterative reconstruction. During reconstruction, the gradient of this energy (the learned score) acts as a data‑driven regularizer that pulls iterates toward the clean‑image manifold while the data fidelity term enforces consistency to the measured k-space data.  Additionally, a standard Tikhonov regularizer is used for the slab profiles, $\mathcal{R}_s(\mathbf S)$, resulting in the following optimization problem:

\begin{equation}
\min_{\mathbf{u},\,\mathbf S}\;\frac{1}{2 \eta^2}
\bigl\lVert\mathcal{A}(\mathbf{u},\mathbf S)-{d}\bigr\rVert_{2}^{2}
\;+\;\lambda_u\,\operatorname{\mathcal{E}_{\theta}}(\mathbf{u})+\lambda_\mathbf S\ \left\lVert \mathbf S - \mathbf S_0 \right\rVert_2^{\,2}\
\end{equation}

\noindent where $\mathcal{E}_{\theta}(\mathbf{u})$ is the energy function modeled by CNN, and $\mathbf S_0$ is the initial guess for the slab profiles. We make use of an alternating minimization scheme where we alternate between (i) a slab combination step that treats the current slab profiles as fixed multiplicative sensitivity maps, and (ii) a slab-profile update step that treats the newly updated image as known. Iterating these two sub-problems until convergence yields joint estimates of the image and slab profiles while avoiding the heavy joint Hessian inversions required by Gauss–Newton used in NPEN. 

The remainder of this section presents the CNN-based regularizer and the alternating optimization framework.

\subsection{CNN-based Regularizer }

For the regularized reconstruction of the 3D slab-combined volume, EPEN employs a learned energy-based regularizer rather than a handcrafted prior or a black-box plug-and-play denoiser. The distinction is important since a plug-and-play denoiser is a pretrained network inserted into an iterative reconstruction in place of a proximal operator, with the underlying regularizer defined only implicitly through the denoiser's behavior \cite{mani2021qmodel}, while the proposed approach uses the energy-model i-MuSE \cite{Chand2024MuSE_TCI}, which defines an explicit scalar energy function whose gradient serves as the regularization term in a well-defined MAP cost. The network is trained to model the prior distribution $p(\mathbf u)$ as: 
\begin{equation}
p_\theta(\mathbf u) = \frac{1}{Z_\theta} \exp\left( {-\mathcal{E_\theta}(\mathbf u)} \right),\\
\end{equation}
where $\theta$ denotes the parameters of the energy function ${\mathcal{E_\theta}(\mathbf u)}$ modeled by the CNN, and $Z_\theta$ is a normalization constant that ensures the prior integrates to unity. The negative log-prior is therefore equal to the energy up to an additive constant; in practice, the constant $Z_{\theta}$ may be ignored, as it does not affect the image reconstruction. The energy term $\mathcal E_{\theta}$ is represented as 
\begin{equation}
\mathcal{E}_{\theta}(\mathbf u) \;=\; \tfrac12\,\bigl\lVert \mathbf{u} - \mathcal D_{\!\theta}(\mathbf{u}) \bigr\rVert_{2}^{2},\\
\end{equation}
where $\mathcal D_{\theta} : \mathbb{C}^N \rightarrow \mathbb{C}^{N}$ is a CNN, and the score $H_{\theta} = -\nabla_{\mathbf u}\mathcal{E}_{\theta}(\mathbf u)$ is computed using the autograd function in PyTorch \cite{paszke2017automatic}.
Because the energy is an explicit, differentiable function, like classical differentiable priors such as Tikhonov, standard optimization algorithms can be used to minimize the posterior, which is the sum of the data-fidelity, likelihood, and prior terms.

\subsubsection{Energy Model Training}

The energy model is trained using denoising score matching (DSM) \cite{li2023learning,vincent2011connection}. While DSM is also used to train generative diffusion models, the network in EPEN is not used generatively. It defines an image prior whose gradient acts as a regularizer within the MAP reconstruction of Eq. (3) and is evaluated only a few times per reconstruction rather than across hundreds of denoising steps.

The training data comprised 3600 artifact-free slices from two healthy volunteers, with 35 minutes total scan time, which was acquired with extended-FOV spacing to suppress slab cross-talk, and the boundary slices were discarded to prevent slab-boundary contamination. 

Since noise-free measurements for DSM training are unavailable, the reference data were denoised using Marchenko-Pastur Principal Component Analysis (MPPCA) \cite{veraart2016denoising}. Further details on training-data preparation and the justification for this denoising step are provided in the Supporting Information section S1.

The energy network is a DRUNet architecture with four downsampling and four upsampling stages, totaling 32 million parameters. Full network architecture details are also provided in the Supporting Information section S2.

\subsection{Alternating optimization framework}

We minimize (3) by alternating between two sub-problems. At every outer iteration, we first update the image while holding the slab profile fixed, and then update the slab profile while holding the image fixed. Each update is guaranteed to lower the joint cost, so the sequence of objective values decreases monotonically until a predefined tolerance is met. 

\subsubsection{Image update}

With $\mathbf  S$ fixed, the forward model becomes linear in $ \mathbf u$:

\begin{equation}
\min_{\mathbf{u}}\;\frac{1}{2 \eta^2}\bigl\lVert \mathcal{A}_{S}\,\mathbf{u}-\mathbf{d} \bigr\rVert_{2}^{2}
\;+\;\lambda_{u}\,\operatorname{\mathcal{E}_{\theta}}(\mathbf{u}), \mathbf{\mathcal{A_S}=PFCS}
\end{equation}

We optimize the above subproblem using the Majorize-Minimize (MM) framework\cite{sun2016majorization}. By replacing the original cost at the $i^{th}$ iteration with a quadratic surrogate function that majorizes (i.e., upper bounds) the original cost, and minimizing the surrogate function to get the next update, the algorithm converges monotonically. At the $i^{th}$ iteration, this surrogate function leads to the closed-form update as follows:
\begin{equation}
\mathbf u_{(t+1)} = \left(\frac{{\mathcal{A}}^{^{H}}_\mathbf S\mathcal{A}_\mathbf S}{\eta^{2}} + L\bm{I} \right)^{-1} \left(\frac{\mathcal{A}^{^{H}}_\mathbf S\mathbf{d}}{\eta^{2}} + L \mathbf u_{(t)} - \mathcal{H}_{\theta}(\mathbf u_{(t)})\right)
\end{equation}
where $L$ is a Lipschitz constant and $\mathcal{H}_\theta = -\nabla_\mathbf u \operatorname{\mathcal{E}_{\theta}}(\mathbf{u}) $ is the negative gradient of the energy, defined as the score. Since direct inversion of the operator $A_S$ is computationally prohibitive, the conjugate gradient (CG) algorithm is used to compute the next iterate $\mathbf u_{(t+1)}$. 

\subsubsection{Slab-profile update}

Fixing $\mathbf u$ turns (3) into the subproblem: 
\begin{equation}
\min_{\mathbf S}\;\frac{1}{2 \eta^2}\bigl\lVert \mathcal{A}_{\mathbf u}\,\mathbf S-\mathbf{d} \bigr\rVert_{2}^{2}\;+\;\lambda_{s}\ \left\lVert \mathbf S - \mathbf S_0 \right\rVert_2^{\,2}\;,\mathbf{ \mathcal{A}_u= PFCu}
\end{equation}
where $\mathbf S_0$ is the initial guess of the slab profiles. Because $\mathbf{A_u}$ is linear in $\mathbf S$, the resulting subproblem is strictly convex and admits the following normal equations for each slab:

\begin{equation}
(\frac{\mathcal{A}_{\mathbf u}^{\!H}\mathcal{A}_{\mathbf u}}{\eta^2} + \lambda_{\mathbf S}\,\bm{I}\bigr)\,\mathbf S = \frac{\mathcal{A}_{\mathbf u}^{\!H}\mathbf{d}}{\eta^2} + \lambda_{\mathbf S}\ \mathbf S_0 
\end{equation}

This normal equation is solved using CG for each slab. After convergence, we enforce physical plausibility by projecting $S$ onto the non-negative orthant, and rescale each slab to eliminate the global scaling ambiguity between $\mathbf u$ and $\mathbf S$.

\subsection{Experiments and Datasets}

To evaluate the performance of the proposed method, we make use of simulations and in-vivo experiments. We first demonstrate the advantage of the proposed regularized reconstruction on synthetic datasets, which uniquely provides a ground truth to facilitate the study. Further experiments show the utility of in-vivo experiments.

\subsubsection{Synthetic Data Details}

To test reconstruction under realistic diffusion contrast and noise, a synthetic single-coil 3D multi-slab dataset was generated from a noisy diffusion-weighted image volume acquired using 2D methods. The same forward acquisition model, including slab profiles and prescribed $k_z$ sampling, was applied so that the measurement-borne noise present in the original 2D DWI is preserved and propagated through the multi-slab physics. Seven slabs were simulated at different acceleration rates using the slab excitation profile extracted from an MRI phantom experiment shown in Figure 2a; details are provided in the Supporting Information section 3. A pseudo-noise-free reference was formed by denoising the noisy 2D DWI with the EPEN denoiser to enable quantitative error reporting. Figure 2b shows this synthetic dataset and the corresponding score maps after denoising using the EPEN denoiser.

A second noise-free synthetic dataset was also utilized for additional experiments, with the details also given in Supporting Information section 3.

\begin{figure}
    \centering
    \includegraphics[width=0.7\linewidth]{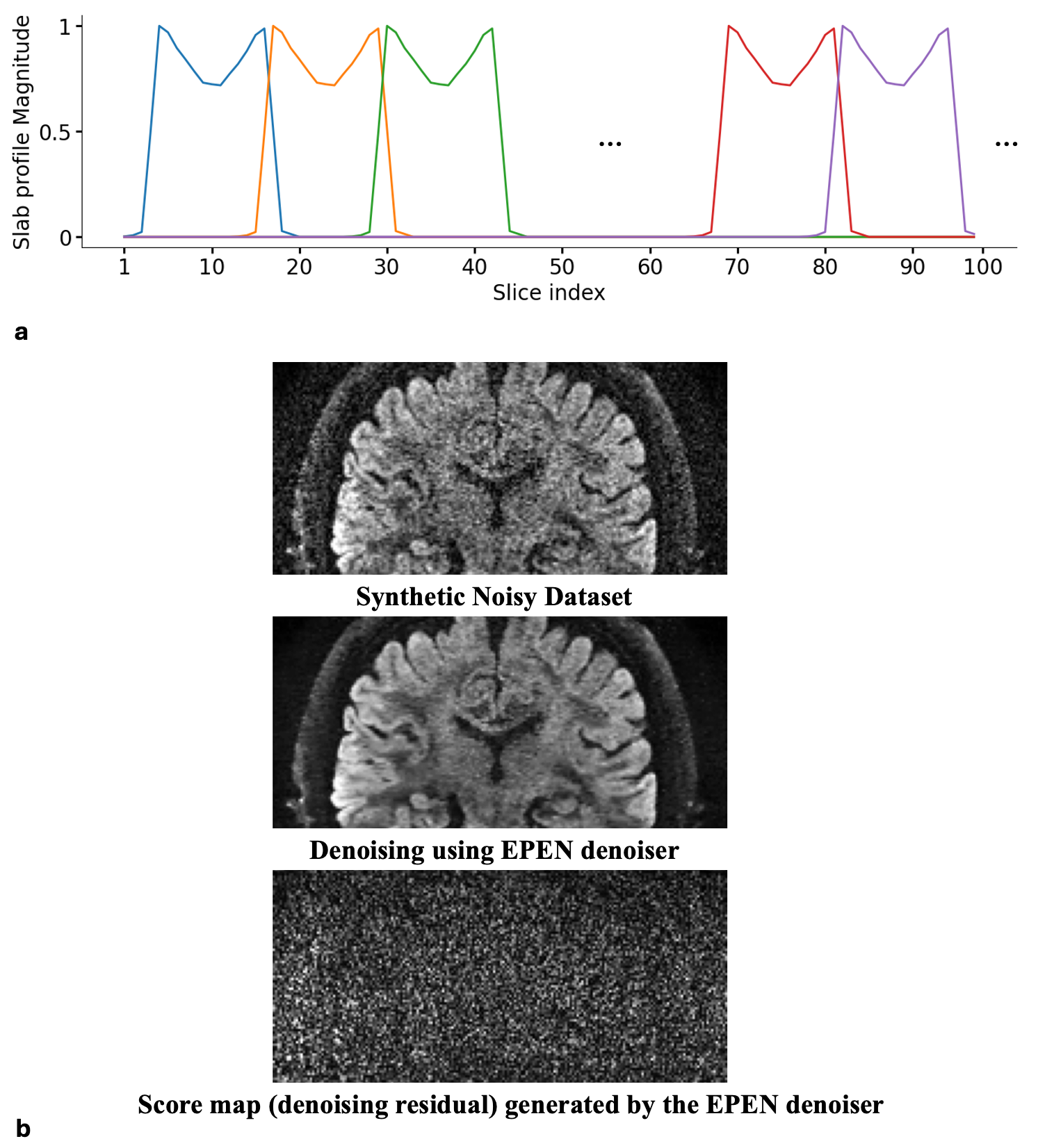}
    \caption{ (a) Slab profiles from the phantom experiment used to generate the synthetic dataset. (b) The synthetic dataset used in this study and example denoising results from the EPEN model. The bottom row shows the learned score (denoising residual), which represents the gradient field the model uses to guide the reconstruction toward the learned data distribution.}
\end{figure}

\subsubsection{In Vivo Data Details}
Seven in-vivo datasets with different acquisition parameters were used to validate the proposed method. All data were acquired under an IRB-approved protocol for human imaging at the University of Iowa. For all datasets, a 3-shot EPI readout was used to fully sample the $k_x-k_y$ plane for $b_0$ image with a partial Fourier (PF) factor of 0.7, and the same EPI readout regime with an in-plane acceleration factor of 3 was used with the same partial Fourier factor for the diffusion-weighted images with b=1000 $s/mm^2$ and b=2000 $s/mm^2$. The full set of acquisition parameters for in-vivo datasets is summarized in Table 1.

Three-dimensional coil sensitivity maps for all slabs were estimated from fully sampled $b_0$ images using sum-of-squares reconstruction. Diffusion-weighted k-space data were subsequently undersampled retrospectively using arbitrary sampling patterns based on a CAIPI sampling grid. For the 14 and 10$k_z$ sampling, certain $k_z$ encodes from the CAIPI grid are removed (Figure S1). TR for all the datasets is set to 2s, and a 2D navigator echo was acquired during each TR and used to correct motion-induced phase errors. The resulting motion-corrected k-space data were used in the proposed reconstruction.

\begin{table*}[h]
\centering
\caption{Acquisition parameters for the in-vivo datasets acquired for this study. Extended-FOV ($\Delta FOV_z$ $-$ slab thickness) is the margin that captures the RF transition bands of the same slab, and the Time/vol is the per-volume acquisition time at the listed number of $k_z$ encodes ($TR \times N_{k_z}$), and the Total acq. is the total scan time for all DW and $b_0$ volumes. The equivalent prospective scan times at any acceleration levels would be reduced by a factor ($N_{k_z,acc} / N_{k_z,full}$). Here S1 represents scanner 1 (GE Premier 3T), and S2 represents scanner 2 (GE MAGNUS 3T).}

\setlength{\tabcolsep}{9pt}
\renewcommand{\arraystretch}{1.12}
\vspace{4pt}

\begin{threeparttable}
\begin{adjustbox}{max width=0.98\textwidth}
\begin{tabular}{@{}ccccccccccccc@{}}
\toprule
Dataset & Scanner & Coil & \# Slabs & $\Delta FOV$ & Extended-FOV & Matrix & TE & $N_{k_z}$ & DW vols & Time/vol & Total acq. & ($N_{k_z,acc}$ , Acc. acq. time)\\
 & & & & (mm) & (mm) & (mm) & (ms) & & & (s) & (min) & ($\#$ , min)\\
\midrule
1 & S1 & 44-ch & 8  & 20 & 6  & $216\!\times\!216$ & 70 & 20 & 36 & 40 & 24.7 & (14,17.3), (10,12.3), (8,9.9)\\
2 & S1 & 44-ch & 8  & 22 & 8  & $240\!\times\!240$ & 70 & 22 & 36 & 44 & 27.1 & (14,17.3), (10,12.3)\\
3 & S2  & 32-ch & 10 & 16 & 2  & $216\!\times\!216$ & 47.5 & 16 & 3 & 32 & 2.2 & (14,1.9), (10,1.4)\\
4 & S2  & 32-ch & 10 & 14 & $0\tnote{*}$ & $216\!\times\!216$ & 48 & 14 & 6 & 28 & 3.3 & (10,2.3)\\
5 & S2  & 32-ch & 10 & 28 & 14 & $216\!\times\!216$ & 48 & 28 & 6 & 56 & 6.5 & (20,4.6)\\
6 & S2  & 32-ch & 10 & 18 & 4  & $216\!\times\!216$ & 49.5 & 18 & $3+3\tnote{**}$ & 36 & 4.2 & (14,3.3)\\
7 & S2  & 32-ch & 10 & 18 & 4.5  & $144\!\times\!144$ & 50.6 & 12 & $3+3\tnote{**}$ & 24 & 2.8 & (9,2.1)\\
\bottomrule
\end{tabular}
\end{adjustbox}

\begin{tablenotes}[flushleft]
\footnotesize
\item[*] Dataset 4 is a non-extended-FOV dataset with 2 mm adjacent slab overlap.
\item[**] Dataset 6 and 7 have 3 b=1000 $s/mm^2$ and 3 b=2000 $s/mm^2$ DW volumes, while in other datasets b=1000 $s/mm^2$.
\end{tablenotes}

\end{threeparttable}
\end{table*}

\section{Results}

We first evaluate the joint reconstruction and slab-combination performance of EPEN on synthetic data with realistic measurement noise, synthetic dataset 2, and compare it against the PEN and NPEN methods at multiple $k_z$ acceleration rates.
Undersampled experiments were performed by retrospectively removing arbitrary $k_z$ encodings, and in each of the accelerated experiments, the slab profile was treated as unknown. The results of all reconstructions are presented in Figure 3. 

\begin{figure}
    \centering
    \includegraphics[width=0.7\linewidth]{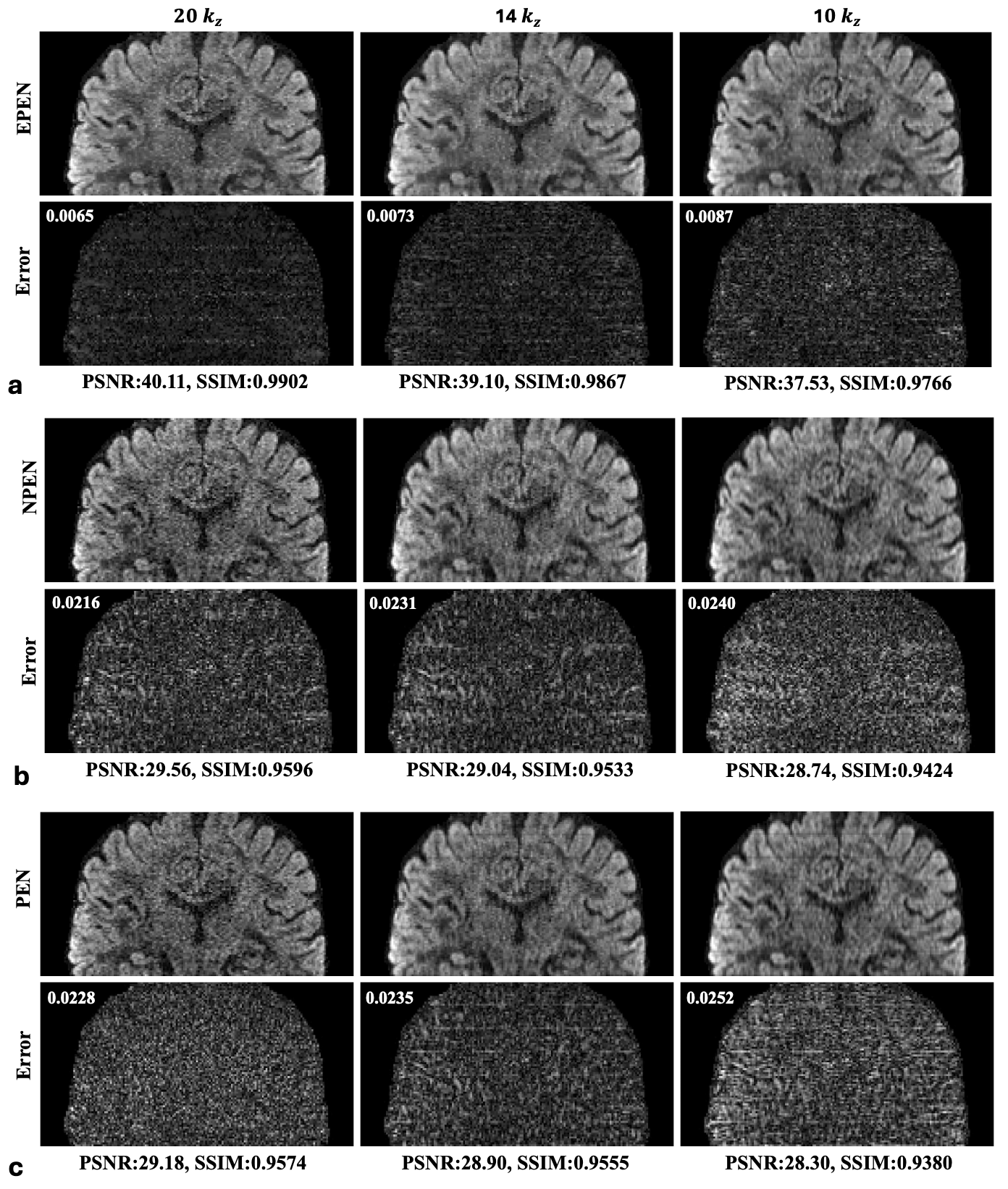}
    \caption{Reconstruction results using the synthetic noisy DWI data for (a) EPEN, (b) NPEN, and (c) PEN methods. The RMSE is shown in the upper left-hand corner of the difference maps (amplified 10 times), computed with respect to the denoised DWI shown in Figure 2b.}
\end{figure}

Specifically, the comparison of the 20 $k_z$ joint reconstruction and slab combination results from all three methods demonstrates reasonable reconstruction across the methods, with the additional observation that EPEN yields a visibly cleaner denoised reconstruction resulting from the prior distribution learning on clean denoised images. Similarly, in the accelerated experiments, the EPEN reconstruction outperforms PEN and NPEN in this single-channel setting. A comparison of the residuals from the accelerated case shows residual slab boundary artifacts, pointing to the noise amplification in the slab profile estimates for NPEN, while PEN assumes fixed slab profiles and lacks any slab profile correction. In the accelerated cases, these methods fail to compensate for the slab profile imperfections. By contrast, EPEN’s integrated energy‑based prior stabilizes the bilinear inversion by stopping the propagation of profile estimation noise back into the image, even when both the slab profile and images are treated as unknowns in the presence of noise. Here, the error maps and the quantitative error metrics are computed with respect to the denoised DWI volume, as described earlier. 

To isolate the contribution of the learned energy prior from confounds such as measurement noise, slab profile estimation error, and coil sensitivity diversity, we additionally performed a controlled single-channel ablation on a noise-free synthetic dataset with the slab profiles treated as fixed and known, which corresponds to solving the optimization in Eq (6). Details of this dataset are provided in Figure S2a. Even in this minimalist setting, the unregularized reconstruction exhibits structured residuals concentrated at slab boundaries that grow with $k_z$ acceleration, while the regularized EPEN reconstruction produces substantially flatter residuals. Quantitative metrics further substantiate the benefit of the learned prior in this setting, as shown in Figure S2b.

Figure 4 shows the multi-channel experiments performed on the in-vivo dataset 1. Here, the EPEN reconstruction clearly shows the denoising on the images in addition to making the reconstruction stable. Thus, EPEN remains robust in the accelerated cases also. NPEN is relatively stable at the 14 k$_z$ sampling rate, but shows slab boundary artifacts at the 10 k$_z$ sampling rate. PEN shows artifacts at both 14 and 10 kz k$_z$ sampling rate as expected, due to the absence of any regularization or slab update. For each of the methods, the RMSE is reported with respect to the respective method's 20 k$_z$ fully sampled case, in the absence of a ground truth.
 
 \begin{figure}
    \centering
    \includegraphics[width=0.7\linewidth]{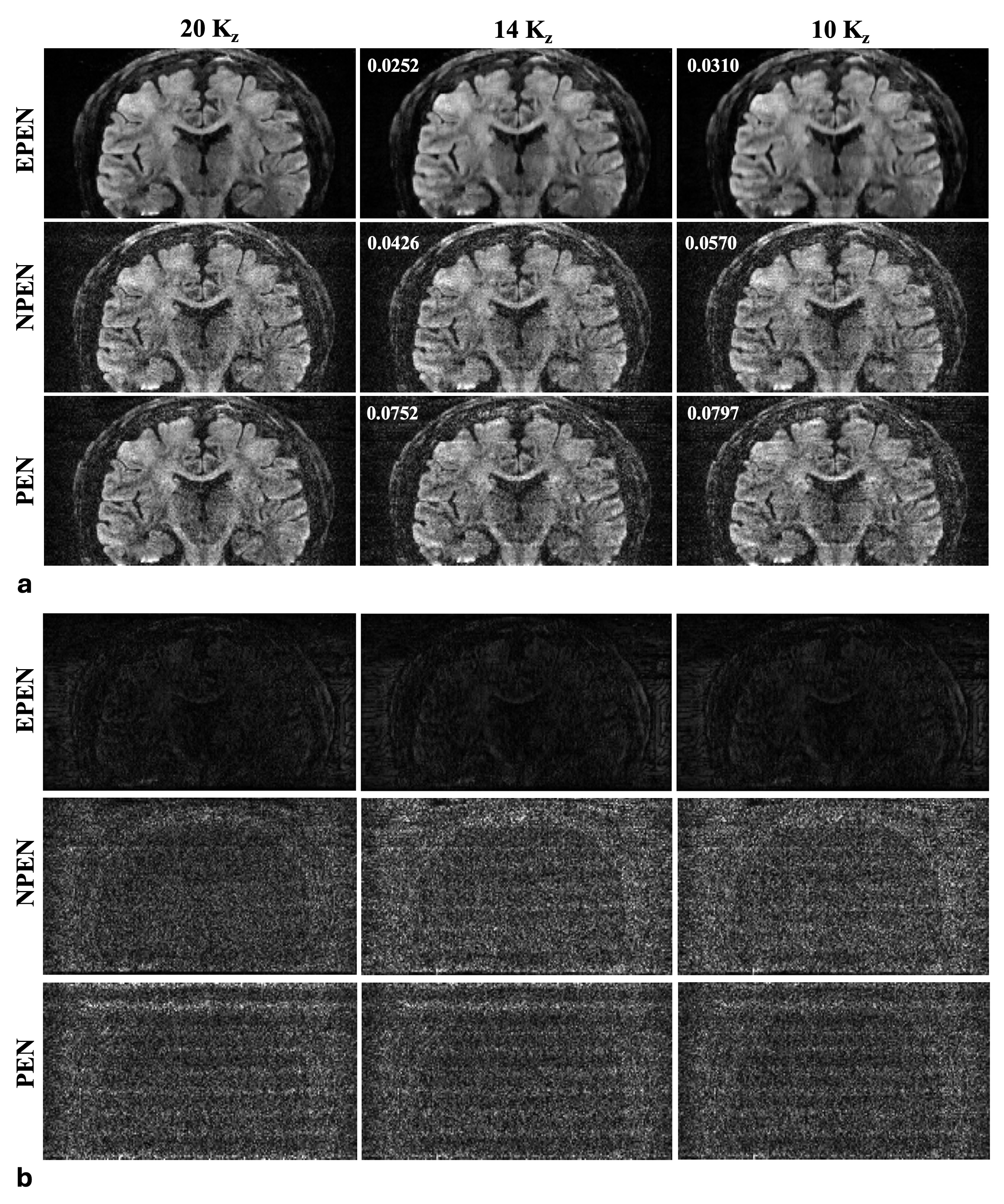}
    \caption{(a) Experiments on in-vivo diffusion-weighted dataset 1. Reconstruction results from the proposed EPEN method are shown at different $k_z$ undersampling rates, in comparison to NPEN and PEN. (b) Score maps corresponding to the reconstructions in (a). Scores can be taken as a proxy for noise or the presence of artifacts.}
\end{figure}

Also, the score maps across EPEN, NPEN, and PEN reconstructions at different $k_z$ undersampling levels are compared in Figure 4b. Score maps provide an additional mechanism for evaluating reconstruction quality. Our learned energy model generates the score whose magnitude reflects the degree of deviation from the clean‑image manifold learned during DSM training. Higher values indicate increased noise or artifact presence or other deviations from the clean manifold. Across all undersampling levels, the EPEN-based reconstructions exhibit consistently lower and more spatially structured score responses, indicating closer adherence to the learned prior distribution of clean images. In contrast, PEN and NPEN reconstructions show markedly elevated and spatially diffuse score magnitudes, indicating noise amplification and residual artifacts. As the sampling density decreases from 20$K_z$ to 10$K_z$, a systematic increase in score magnitude is observed for all methods, reflecting reduced data consistency and reconstruction accuracy. However, even at the high undersampling, EPEN maintains comparatively low score responses, suggesting more effective constrained reconstruction to the clean-image manifold and a superior robustness to the through-plane undersampling. To quantify how this robustness varies with slice position, we additionally computed the slice-wise RMSE and slice-wise SSIM, both relative to each method's own 20 $k_z$ reference, and plotted them against z-position (Figure S3). It is shown that the EPEN maintains consistent image quality across both central and boundary positions along the slice direction.

In these experiments, the 20 $k_z$ fully sampled acquisition serves as a methodological reference for quantitative error comparison rather than as a recommended operating point; its $40\%$ oversampling ensures that the reference image is free of residual slab-related artifacts. A complementary experiment using a less-oversampled reference with 16 $k_z$, $\approx14\%$ oversampling, dataset 3, is reported in Figure S4. EPEN consistently outperforms NPEN at 14 and 10 $k_z$ under this reference, also confirming that the reconstruction-quality benefits of EPEN are not contingent on the 20 $k_z$ reference used in the main experiments.

\begin{figure}
    \centering
    \includegraphics[width=0.7\linewidth]{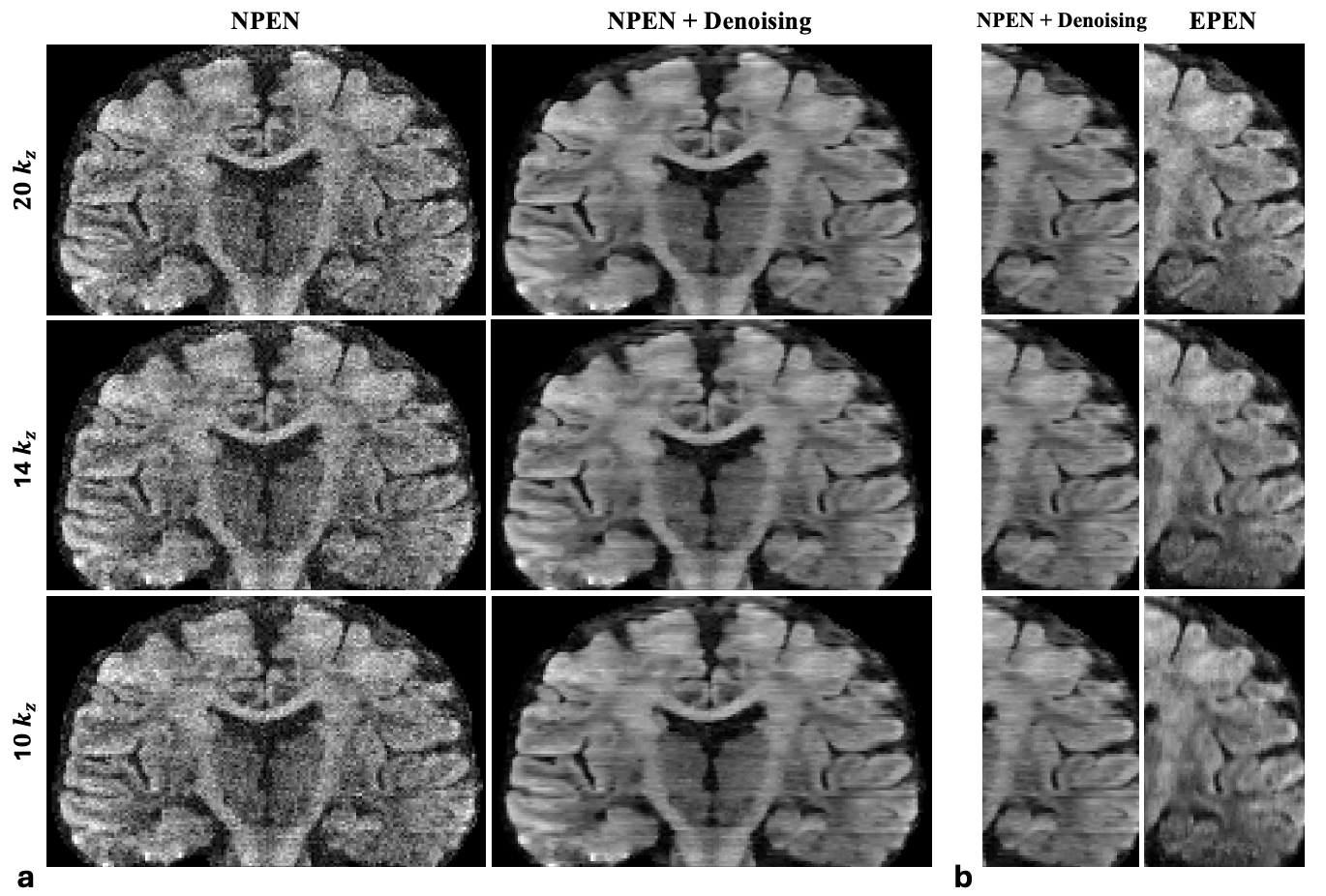}
    \caption{ (a): Post-denoising of NPEN reconstruction using the learned CNN-based denoiser. Although the images are denoised, the artifacts cannot be removed in this post denoising. (b): A selected region of the denoised NPEN reconstruction and its corresponding region from the EPEN reconstruction.}
\end{figure}

We further evaluated whether post-reconstruction denoising of the PEN and NPEN reconstructions could achieve performance comparable to EPEN. Figure 5 presents this comparison, in which the same learned CNN denoiser was applied to the NPEN results as a post-processing step. The results of applying the same procedure to PEN reconstruction is provided in Figure S5. Although post-denoising reduced noise in the PEN and NPEN reconstructions, slab artifacts could not be suppressed; instead, they became more pronounced and resulted in blurred images. This experiment further demonstrates that the learned prior incorporated into the iterative reconstruction is more effective to jointly stabilize the image and slab‑profile updates than is achievable through post‑processing.

 Figure 6 shows the results of a tensor fitting experiment \cite{o2011introduction} of in-vivo Dataset 1 at the three k$_z$ sampling rates.  DWI volumes from 36 directions and the one b0 volume were reconstructed individually using EPEN, NPEN, and PEN, and the resulting images were used for tensor fitting using the DIPY toolbox \cite{garyfallidis2014dipy}. The fractional anisotropy (FA), mean diffusivity (MD), and the color-coded FA maps are displayed for all the acceleration factors. The FA maps from the EPEN results are less noisy than expected. Residual slab artifacts from NPEN and PEN are visibly apparent in the MD maps, and are noticeable on the FA maps as well. Additionally, we performed a comparison of fiber tractography results across EPEN reconstruction using DSI Studio \cite{yeh2025dsi}, where the slice direction maintains consistent fiber tract quality across different sampling rates, with no observable degradation (Figure S6).

\begin{figure}[h]
    \centering
    \includegraphics[width=0.6\linewidth]{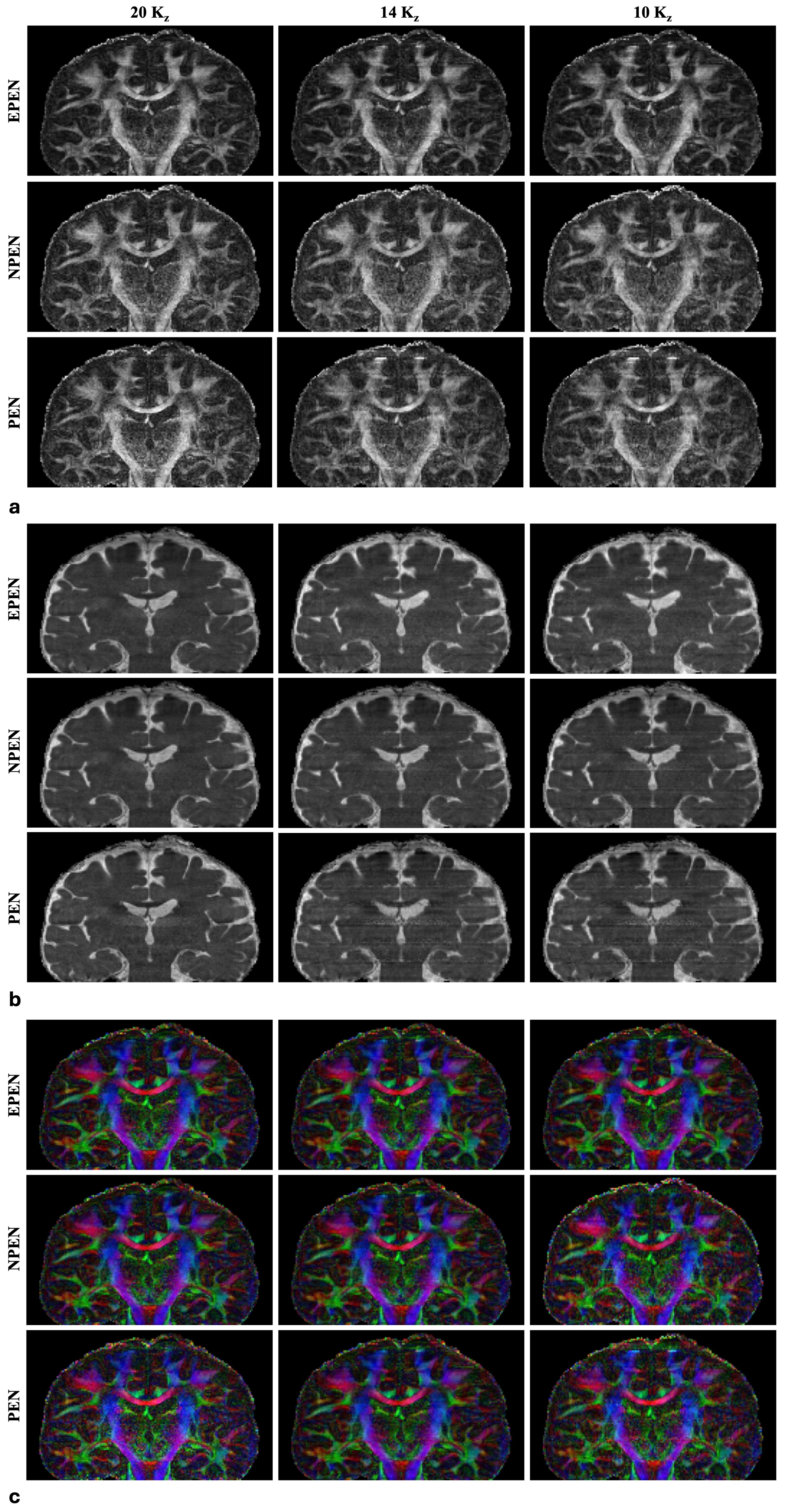}
    \caption{The parameter maps estimated from DTI model fitting using in-vivo dataset 1. The fractional anisotropy (FA) (a), mean diffusivity (MD) (b), and color-coded FA (c) maps are shown from the three methods. EPEN produces FA maps with reduced noise relative to the other reconstructions, while residual slab artifacts present in NPEN and PEN reconstructions propagate into the tensor estimates and affect the corresponding maps.}
\end{figure}

Figure 7 shows the results of similar experiments from the in-vivo Dataset 2. The same trends are observed in this data, where EPEN provides robust reconstruction in the presence of acceleration. The NPEN reconstruction is shown for comparison.

\begin{figure}
    \centering
    \includegraphics[width=1\linewidth]{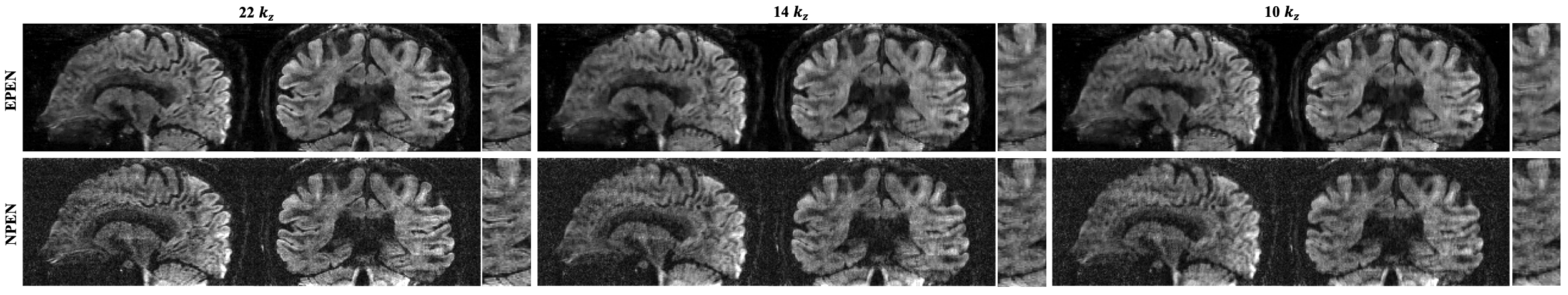}
    \caption{ The reconstruction results for the in-vivo dataset 2 at the different k$_z$ sampling rates. EPEN reconstructions are shown on the top row, and NPEN reconstructions are shown on the bottom row.}
\end{figure}

We further extended the acceleration by employing a more aggressive 8$k_z$ sampling rate. Under this condition, we compared the proposed EPEN method to NPEN across in-vivo datasets 1 and 2. The results, presented in Figure 8, demonstrate that EPEN reconstruction outperforms NPEN reconstruction in terms of image reconstruction quality and slab combination artifact removal as expected, although at the expense of blurring in the resulting images from the high rate of acceleration. Additional q-space regularization is helpful in such cases to improve the reconstruction quality \cite{Lee2026}.

\begin{figure}
    \centering
    \includegraphics[width=0.6\linewidth]{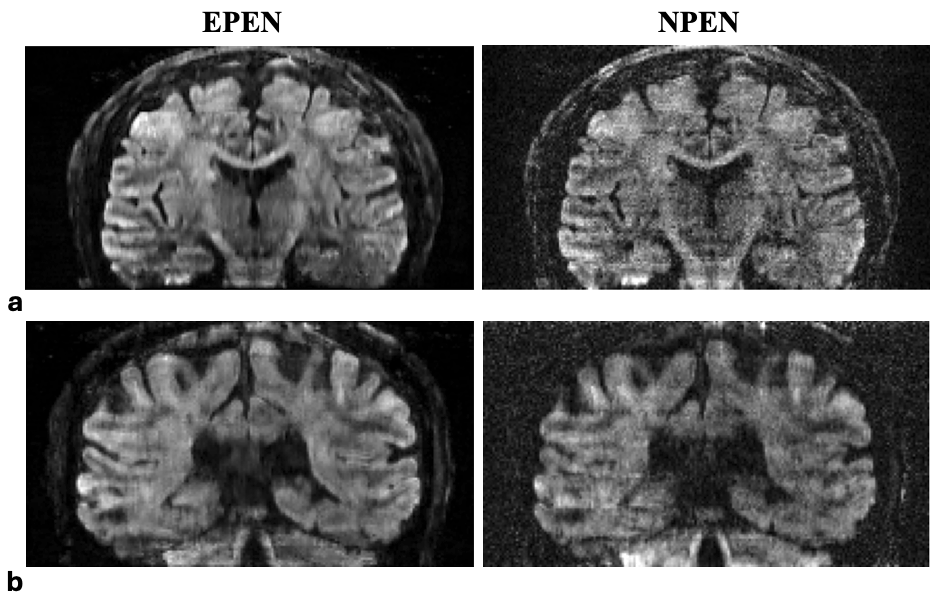}
    \caption{ The reconstruction results for the in-vivo dataset 1 (a) and dataset 2 (b) for the case with 8 $k_z$.}
\end{figure}

Figure 9 shows the results of experiments performed on dataset 4. This dataset was acquired with the traditional PEN-style acquisition, where the phase-encoded FOV is kept equal to the slab thickness using 14 $k_z$ encodes. Also, this dataset was retrospectively undersampled to generate 10 $k_z$ encodes and was reconstructed using EPEN, NPEN, and PEN. Here, NPEN and PEN relied on calibration data extracted from dataset 5, whereas EPEN was self-calibrated. A comparison of the reconstructions shows that, in the classical phase-encoding setting, EPEN can maintain a superior performance compared to the NPEN and PEN reconstructions. The results of the accelerated experiments on dataset 5 are provided in Figure S7, and the DTI analysis for datasets 4 and 5 is shown in Figure S8.

\begin{figure}
    \centering
    \includegraphics[width=0.6\linewidth]{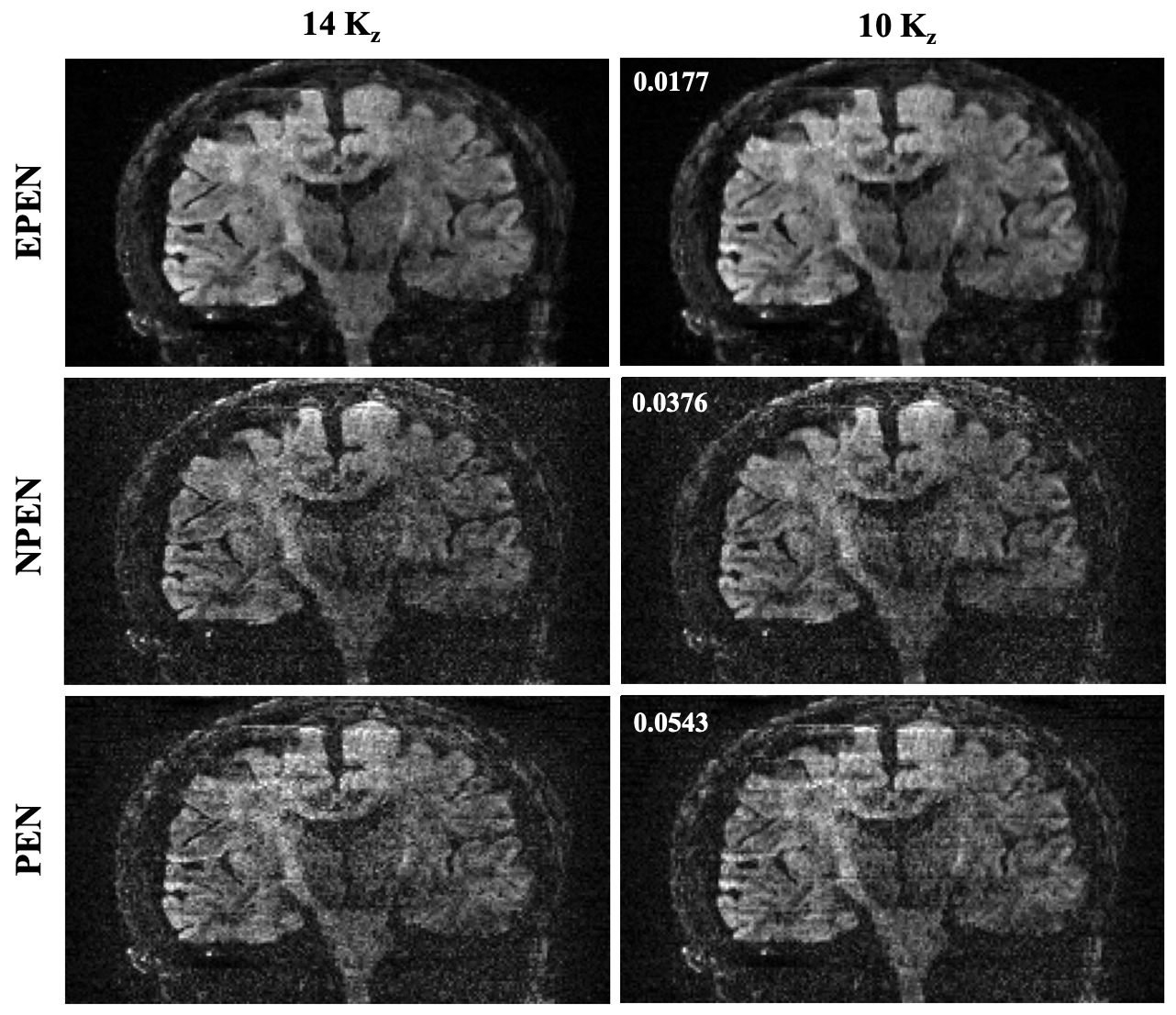}
    \caption{ The reconstruction results for the in-vivo dataset 4 with 14 $k_z$, and 10 $k_z$.}
\end{figure}

To assess the sensitivity of the learned prior to the amount of training data, we trained the energy model with approximately 1600, 2600, and 3600 slices and evaluated each as a denoiser on a held-out validation set. Denoising performance was stable across this range, with PSNR varying by less than 0.6 and SSIM degrading only gradually (Table S1), indicating that the prior is data-efficient and not strongly dependent on the training-set size within this regime. Additionally, the generalizability of the trained energy model across acquisition parameters not represented in the training was evaluated for EPEN on two in-vivo datasets 6 and 7 acquired at 1 mm and 1.5 mm isotropic resolution and at b=1000 and b=2000 $s/mm^2$ (Figure S9). All reconstructions used the same energy-model checkpoint as the main experiments. The reconstruction quality is preserved across all resolution/b-value combinations, indicating that the learned prior captures broadly applicable image statistics rather than acquisition-specific features.

\section{Discussion}

In this work, we presented EPEN, an energy–regularized slab profile encoding framework for 3D multi-slab dMRI reconstruction that jointly estimates the underlying 3D image and slab excitation profiles within a maximum a posteriori formulation. By integrating a learned deep energy prior within a physics-based forward model and a majorize-minimize optimization strategy, EPEN extends prior slab profile encoding methods by improving both reconstruction fidelity and robustness under accelerated acquisition conditions.

Classical PEN treats slab boundary artifacts as a linear unmixing problem with predetermined slab profiles, which performs well under full sampling but degrades when these profiles are imperfect, or encoding becomes ill-conditioned. NPEN improves robustness through joint image- slab profile estimation, but relies on handcrafted regularization and Gauss–Newton updates sensitive to noise-amplified slab profile errors under acceleration. EPEN generalizes NPEN by replacing the handcrafted regularizer with a data-driven deep energy prior trained via DSM, which models a richer class of image statistics within the same bilinear physical model, while adopting a self-calibrated approach in which slab profiles are estimated directly from the data. The results demonstrate consistently lower slab-boundary artifacts and improved image fidelity compared with both PEN and NPEN, particularly under $k_z$ undersampling.

EPEN shares with CPEN the goal of integrating learned components into slab profile encoding reconstruction, and both jointly estimate the image and slab profiles. However, the approaches differ substantially, where CPEN trains a CNN end-to-end as a learned update operator replacing the conjugate-gradient solver within a Gauss–Newton iteration, while EPEN embeds an energy-based prior into an explicit MAP optimization with majorize–minimize updates and monotonic-descent guarantees, as mentioned earlier. The two methods also target different regimes where CPEN addresses oversampled acquisitions where the slab-mixing inversion is well-conditioned, whereas EPEN targets the more ill-conditioned $k_z$-undersampled regime via the extended-FOV strategy, which has no counterpart in CPEN. Because EPEN's prior models the distribution of clean diffusion-weighted images rather than a specific corruption-to-clean mapping, it generalizes across acquisition configurations using a single trained model. A direct head-to-head comparison is not straightforward given these differences, but it remains an interesting direction for future work.

Classical PEN implementations favor using thin slabs with more slices than slabs to satisfy algebraic constraints that stabilize the inversion. However, this is not ideal for volumetric excitation, where thicker slabs can deliver higher SNR, especially when pushing towards the signal-starved domains of high spatial resolution or high b-values. In EPEN, we explicitly operate in a thick-slab regime, where we decouple slab thickness from the phase‑encoded FOV in the slice direction and realize an extended FOV by skipping a subset of k$_z$ lines, keeping the number of encodes and the scan time fixed. This strategy converts coherent fold‑over into more distributed aliasing that is easier to resolve in a joint reconstruction, and it eliminates the need for thin slabs solely for inversion stability. With thicker slabs, the TR can be maintained close to 2 seconds per $k_z$ encode, which is critical for applications requiring the acquisition of a large number of volumes (eg: fiber tractography, advanced microstructure modeling, etc.) Also, the reduced $k_z$ sampling pattern retains a portion of the high-$k_z$ samples in an asymmetric manner, rather than fully truncating high-frequency information,  thereby reducing the severity of any associated loss of through-plane resolution. EPEN mitigates the resulting blurring and artifacts by stabilizing the inverse problem through coil sensitivity encoding and a learned energy prior, promoting structurally consistent solutions under incomplete sampling. This behavior is consistent with standard regularized partial Fourier reconstruction and is supported by the preservation of structure in RMSE, SSIM, tractography, and DTI analyses. We note that further improvements may be achievable through optimized $k_z$ sampling strategies that better preserve high-frequency information. In addition, joint reconstruction approaches that exploit correlations across diffusion directions could further mitigate the effects of incomplete $k_z$ sampling. These directions are beyond the scope of the present work and will be explored in future studies.

As mentioned, the central contribution of EPEN is the integration of a CNN-based energy model into a MAP reconstruction framework, which, unlike unrolled network approaches, provides a well-defined objective function and principled incorporation into an MM-based optimization scheme with monotonic descent guarantees, while avoiding the contraction constraints required for convergence in traditional plug-and-play methods. The score-based prior effectively suppresses noise amplification and residual aliasing while preserving fine structural details that are typically degraded by quadratic or total-variation regularization, since capturing higher-order spatial dependencies not easily expressible using conventional regularization. Importantly, these improvements persist even when PEN and NPEN reconstructions are post-processed using the same trained denoiser, confirming that the performance gains arise from joint reconstruction regularization rather than post hoc denoising alone.

Also, it is shown using simulation and in-vivo experiments that EPEN exhibits improved robustness across a range of acquisition conditions. Under short-TR conditions, where slab profile distortion is most severe, EPEN maintains stable image quality and accurate slab profile estimates, while the performance of PEN degrades substantially. This robustness can be attributed to the joint estimation framework combined with the strong statistical constraints imposed by the learned energy prior.

For diffusion MRI applications, residual slab-boundary artifacts can propagate into diffusion tensor and microstructural parameter estimates, leading to spatially varying biases in FA, MD, and orientation estimates. The results demonstrate that EPEN not only suppresses visible slab artifacts but also significantly reduces bias in derived diffusion metrics, particularly in regions adjacent to slab boundaries where PEN and NPEN exhibit residual errors. This highlights the importance of addressing slab artifacts at the reconstruction level rather than attempting to correct them in post-processing.

However, although computationally feasible, EPEN remains more expensive than PEN and NPEN, particularly for high-resolution 3D datasets. The additional computational cost relative to NPEN arises primarily from repeated score evaluations through the energy model during the majorize–minimize image update. Full timing details are provided in the Supporting Information section S2.

Additionally, diffusion-gradient-induced geometric distortions in our 3D multi-slab DW-EPI acquisition are predominantly in-plane, and the slab-selective direction is comparatively less affected. However, residual through-plane distortions may become more pronounced at higher b-values or in systems with less linear gradient performance. In such regimes, incorporating geometric distortion models into the reconstruction framework represents a potential extension for improved robustness.

Future work will focus on several extensions of the proposed framework. Incorporation of diffusion-aware priors and multi-contrast energy models may further enhance reconstruction accuracy for advanced microstructural modeling. Extension of EPEN to simultaneous multi-slab acquisitions may enable improved performance at ultra-high field. Finally, hybrid formulations that combine energy-based priors with diffusion-based score models may provide further gains in robustness and expressiveness.

\section{Conclusion}

In summary, EPEN provides a unified, physics-guided, and statistically principled framework for 3D multi-slab dMRI reconstruction by jointly estimating slab profiles and incorporating an energy-based image prior within a provably convergent MAP optimization scheme. By addressing both the physical origin of slab-boundary artifacts and the statistical structure of anatomical images, EPEN achieves superior artifact suppression, improved quantitative accuracy, and enhanced robustness across a wide range of acquisition conditions compared with existing slab profile encoding methods. These results establish EPEN as a general and extensible reconstruction framework for high-resolution 3D multi-slab diffusion imaging.

\section{Data and Code Availability Statement}

The reconstruction code, pretrained energy-model weights, and a representative example dataset will be available upon reasonable request to the corresponding author.

\section{Acknowledgments}

This research was supported in part by the National Institute of Biomedical Imaging and Bioengineering under Grant Nos. 1S10OD025025-01, 1S10OD030220-01, and 1S10RR028821-01;  and the National Institute of Health under Grant Nos. R01EB031169, and EB031169-02S1. The authors also acknowledge the support of the University of Iowa for providing the facilities and resources necessary for this study.

\bibliography{sample}

\newpage
\setcounter{figure}{0}
\renewcommand{\thefigure}{S\arabic{figure}}
\setcounter{table}{0}
\renewcommand{\thetable}{S\arabic{table}}
\setcounter{section}{0}
\renewcommand{\thesection}{S\arabic{section}}

\section*{Supporting information}

\section{Energy Model Training Details}

The energy model in EPEN is trained via denoising score matching (DSM) \cite{li2023learning,vincent2011connection} over a bounded noise range sufficient for regularization, rather than the full noise schedule required for generative diffusion models. A diffusion model must learn the score from near-zero noise, up to pure Gaussian noise, because generation proceeds by progressively denoising pure noise through every scale, requiring tens to hundreds of sequential network passes at inference. Because EPEN uses the prior solely as a MAP regularizer and never for generative sampling, the bounded training range and small number of score evaluations per reconstruction mean that EPEN retains the computational and training-data footprint of a standard CNN-based regularized reconstruction rather than that of a diffusion model.

Training references were obtained by spacing slabs so that each slab's extended-FOV margin met that of the neighboring slab, suppressing slab cross-talk at acquisition. Central slices were extracted from each slab where the RF excitation profile is approximately uniform, and the boundary slices were discarded. This combination of physical inter-slab gap and boundary-slice removal yields well-separated, artifact-free slices. Since noise-free measurements are unavailable, references were denoised using Marchenko-Pastur Principal Component Analysis (MPPCA) \cite{veraart2016denoising}, chosen for its established role in dMRI preprocessing and its parameter-free, noise-distribution-agnostic operation. 

For score-based models, training on denoised rather than noisy references is important since the DSM objective would otherwise learn the score of the noisy distribution rather than the clean-image prior, a limitation documented in recent score-model literature \cite{aali2023solving}. Because MPPCA is a fixed classical denoiser, the achievable reconstruction quality is bounded by the quality of its output; learned or self-supervised alternatives such as SURE-Score \cite{aali2023solving}, and Noise2Noise \cite{lehtinen2018noise2noise} are not constrained by MPPCA's assumptions and could, in principle, yield higher-quality references, at the cost of paired acquisitions or explicit noise modeling. Optimizing this denoising step is a direction for future work.

\section{Network Architecture and Computational Details}
The energy network is a DRUNet comprising four downsampling and four upsampling stages with 64, 128, 256, and 512 feature channels, respectively, each incorporating four residual blocks and connected via skip connections. The network contains 32 million parameters, which is compact relative to the 50–114 million parameters typically used in denoising diffusion models with comparable architectures \cite{ho2020denoising}. Complex-valued inputs are handled as two separate channels \cite{Chand2024MuSE_TCI}.
In terms of computational cost, EPEN reconstructed a single diffusion-weighted volume in approximately 5.3 minutes using 3 outer alternating iterations with 20 iterations per image-update sub-problem, on an NVIDIA A100 40 GB GPU. For comparison, NPEN used 10 outer Gauss–Newton iterations with 50 inner CG iterations, requiring approximately 4.2 minutes per volume on the same hardware. All timings were measured using a representative in-vivo dataset (Dataset 1, 14 $k_z$).

\section{Synthetic Dataset Generation}
To simulate the 3D multi-slab acquisition, the slab excitation profile was extracted from an MRI phantom experiment, where a slab thickness of 14 mm and a slab FOV of 20 mm ($40\%$ oversampling) was prescribed, and 20 $k_z$ encodes were employed to sample the FOV fully. The excitation and refocusing pulses were designed using the Shinnar-LeRoux algorithm \cite{pauly1991parameter} with a time-bandwidth product of 12; the durations of the excitation and refocusing pulses were 5.1 and 17.9 ms, respectively. 
The slab profile of the central slab from the phantom experiment was estimated by dividing the single-slab image by the sum-of-squares and was treated as a representative template tiled axially across all slab positions to generate slabs with $20\%$ extended-FOV margin. Three cases were studied: a fully sampled case (20 $k_z$) and two accelerated cases (14 $k_z$ and 10 $k_z$).

\newpage
\begin{table}
\centering
\caption{Denoising performance of the EPEN energy model trained with different numbers of training slices, evaluated on a held-out validation set. Metrics are computed against the MPPCA-denoised references. All trained models substantially improve over the noisy input, and performance degrades only gradually as the training set is reduced.}
\vspace{4pt}

\begin{tabular}{lcc}
\toprule
\textbf{Training samples} & \textbf{PSNR} & \textbf{SSIM} \\
\midrule
Noisy input (baseline)      & $20.84 \pm 0.25$ & $0.2049 \pm 0.06$ \\
$\sim$1600 (30\%)           & $31.65 \pm 1.26$ & $0.6570 \pm 0.04$ \\
$\sim$2600 (50\%)           & $31.73 \pm 1.34$ & $0.6714 \pm 0.04$ \\
$\sim$3600 (70\%, used in paper) & $32.22 \pm 1.61$ & $0.7208 \pm 0.03$ \\
\bottomrule
\end{tabular}
\end{table}

\begin{figure}
    \centering   
    \includegraphics[width=0.7\linewidth]{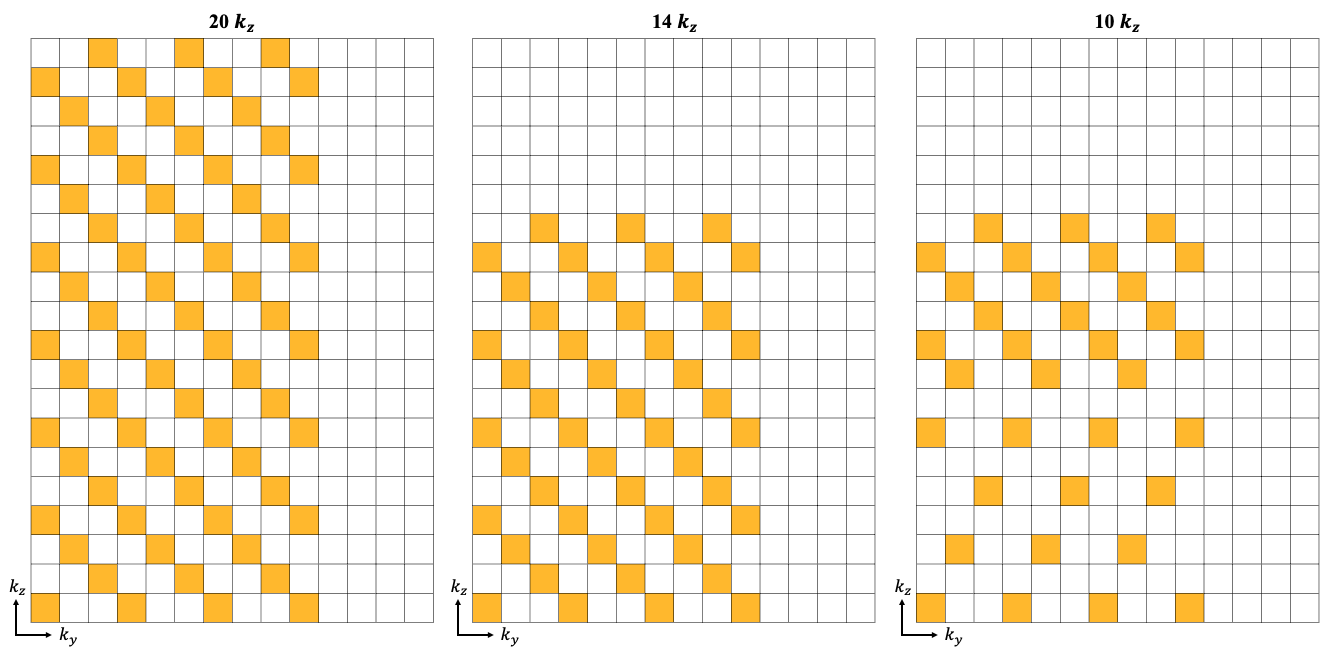}
    \caption{The sampling pattern used for the fully sampled 3D slab acquisition is shown in (a) for the 20 k$_z$ case. Here, a 2D CAIPI sampling pattern is achieved using a shot-selective method with 3 EPI shots. For the accelerated cases, retrospective under-sampling was performed by removing selected k$_z$ encodes from the fully sampled case. (b) shows the 14 k$_z$ case and (c) shows the 10 k$_z$ case.  }
\end{figure}

 \begin{figure}
    \centering   
    \includegraphics[width=0.5\linewidth]{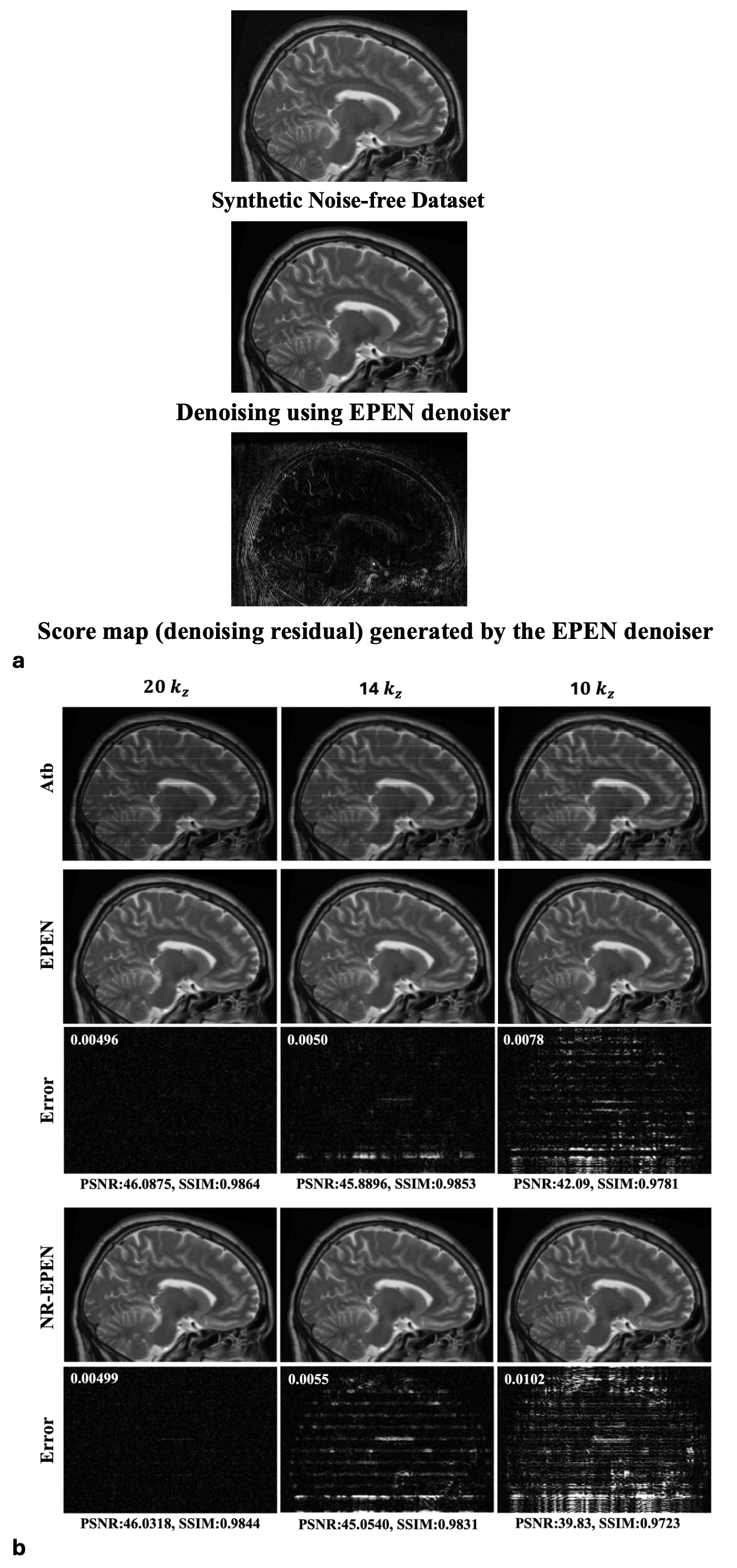}
    \caption{ (a) The synthetic noise-free dataset generated using a high-resolution T2-weighted image from the MNI-Colin27 multimodal brain atlas and example denoising results from the EPEN model. To simulate the 3D multi-slab acquisition, the same procedure to generate the DWI synthetic dataset was applied to generate 12 slabs with $20\%$ extended-FOV margin. The energy model, trained only on diffusion data, generalizes well to T2‑weighted images. The bottom row shows the learned score (denoising residual), which represents the gradient field the model uses to guide the reconstruction toward the learned data distribution. (b) The proposed regularized slab-combination demonstrated on simulated noise-free data at (i) 40\% oversampling (20$k_z$), (ii) no oversampling (14 $k_z$), and (iii) 30\% under-sampling (10 $k_z$), and its comparison to an ablated method with no regularization. This controlled setting, which is selected to be noise-free, single-channel, known slab profiles, isolates the contribution of the learned energy prior from confounds such as measurement noise, profile-estimation error, and coil-sensitivity diversity. The error maps are amplified by a factor of 20 to reveal residual artifacts, and the RMSE values are shown in the left-hand corner. The regularized reconstruction is shown to provide robust reconstructions compared to the non-regularized version, with known slab profiles.}
\end{figure}

\begin{figure}
    \centering
    \includegraphics[width=0.52\linewidth]{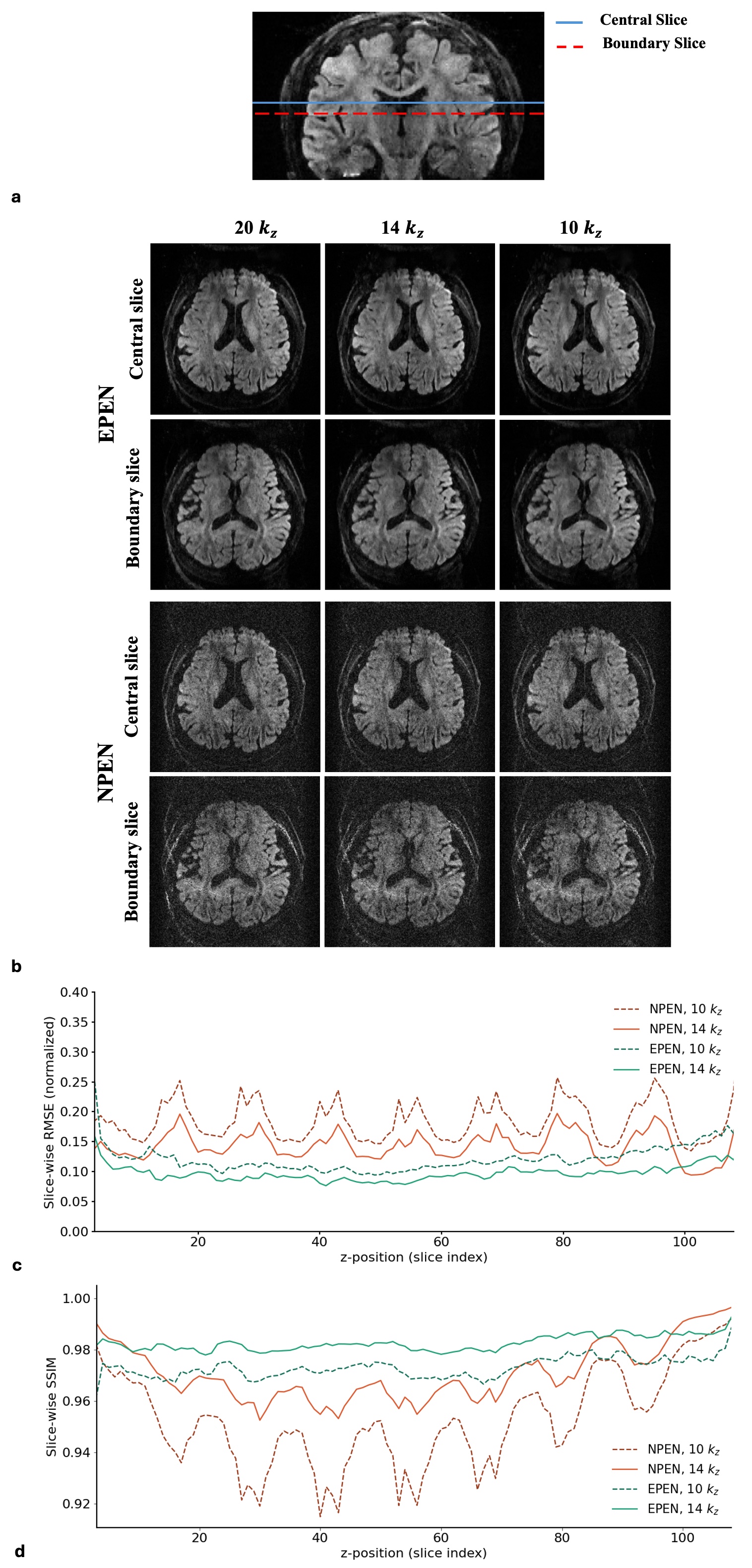}
    \caption{Comparison of image-quality degradation between selected slab-central and slab-boundary slices under $k_z$ undersampling. (a) Coronal localizer with horizontal markers indicating the central (solid) and boundary (dashed) slice positions used for the comparison. (b) Axial views of representative central and boundary slices reconstructed using EPEN and NPEN at 20 $k_z$, 14 $k_z$, and 10 $k_z$. NPEN shows pronounced intensity inhomogeneity and residual structure concentrated at boundary positions that worsen with acceleration, while central slices are comparatively preserved, whereas EPEN maintains consistent image quality across both central and boundary positions. (c) Quantitative slice-wise RMSE plotted against z-position, computed relative to each method's own 20 $k_z$ reference. NPEN exhibits clear peaks at every slab interface that grow with acceleration, while EPEN remains nearly flat, confirming that the joint regularized formulation neutralizes the boundary-specific vulnerability. (d) Quantitative Slice-wise SSIM plotted against z-position, computed relative to each method’s own 20 $k_z$ reference. NPEN curves exhibit pronounced periodic SSIM drops at every slab boundary that deepen with acceleration and additionally show reduced SSIM in central slab regions, whereas EPEN curves remain near unity across the entire z-axis at both 14 $k_z$ and 10 $k_z$, with only minimal residual dips at slab interfaces.}
\end{figure}

\begin{figure*}[t!]
    \centering
    \includegraphics[width=\linewidth]{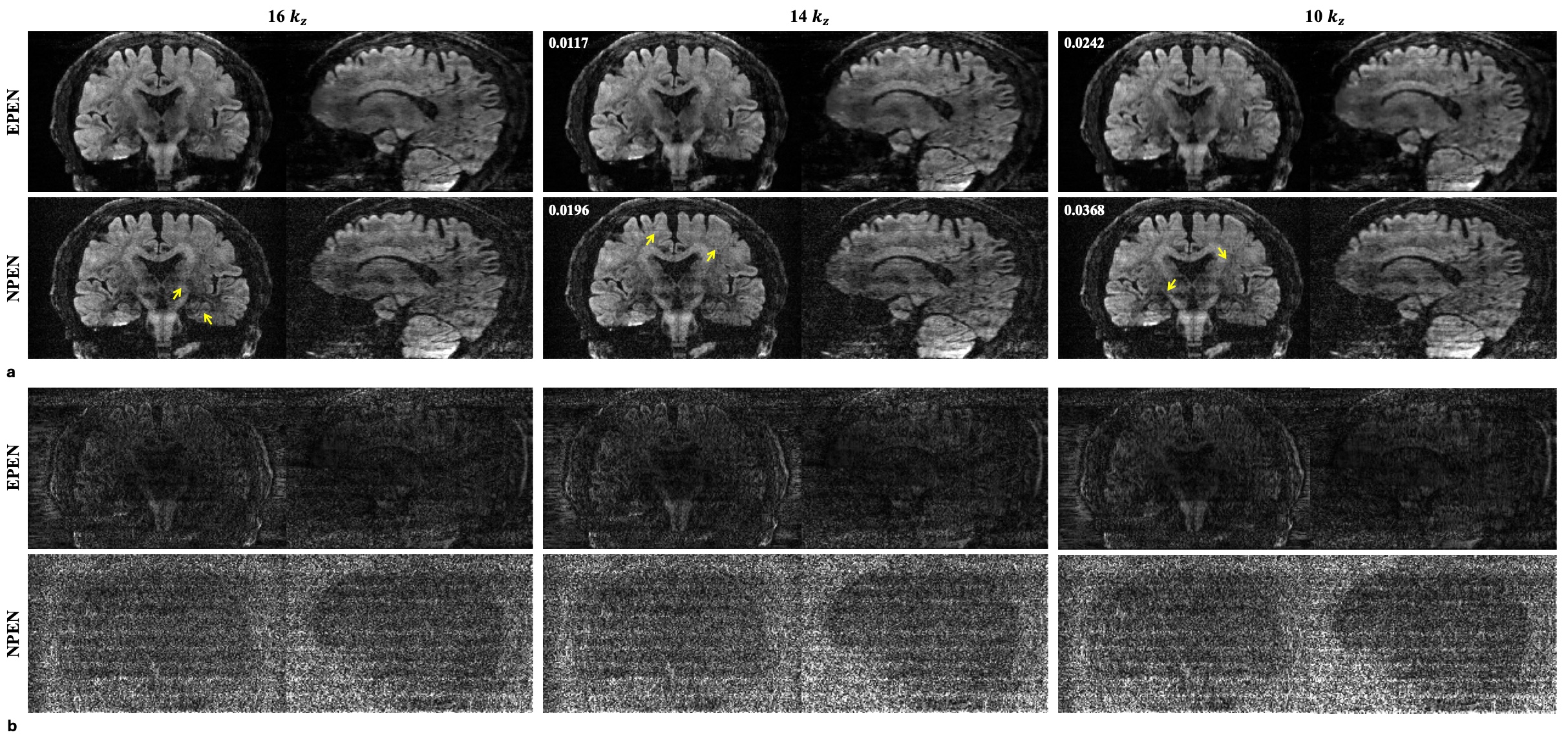}
    \caption{Reconstruction comparison using dataset acquired with 16 $k_z$. (a) Sagittal and coronal views of reconstructions from EPEN and NPEN at the two undersampled $k_z$. The RMSE values are shown in the upper left-hand corner of the reconstructed images, and the yellow arrows indicate residual slab-boundary artifacts, showing that these artifacts become more pronounced with increasing undersampling in NPEN, whereas EPEN continues to effectively suppress them even at higher undersampling rates; (b) Corresponding score maps. The 16 $k_z$ acquisition ($\sim 14\%$ oversampling) provides a more practical reference baseline than the 20 $k_z$ reference used in the main experiments. Across the comparisons, EPEN outperforms NPEN at both 14 and 10 undersampled $k_z$, as observed in the main 20 $k_z$ experiments, which confirms that the EPEN advantage does not depend on the choice of reference baseline.}
\end{figure*}

\begin{figure*}[t!]
    \centering   
    \includegraphics[width=0.7\linewidth]{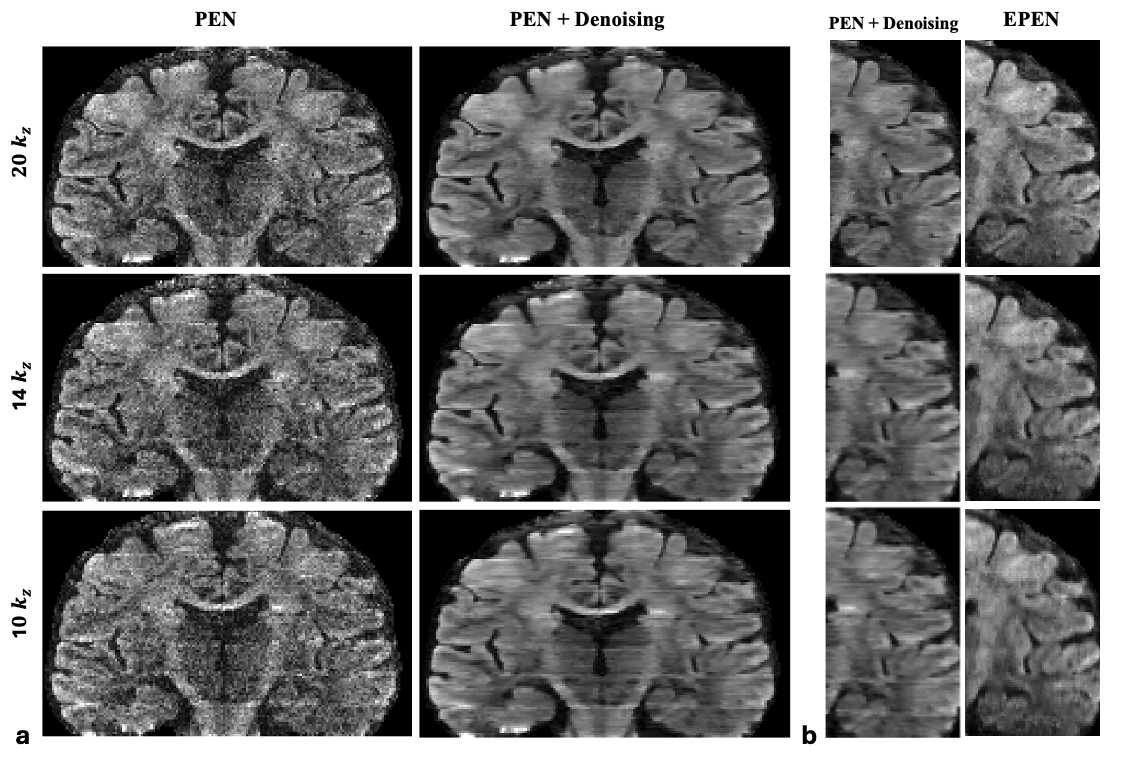}
    \caption{Post-denoising of PEN reconstruction using the learned CNN-based denoiser. Although the images are denoised, the artifacts cannot be removed in this post denoising (a). A selected region of the denoised PEN reconstruction and its corresponding region from the EPEN reconstruction (b).}
    \label{fig:sequence}
\end{figure*}

\begin{figure*}[t!]
    \centering   
    \includegraphics[width=0.7\linewidth]{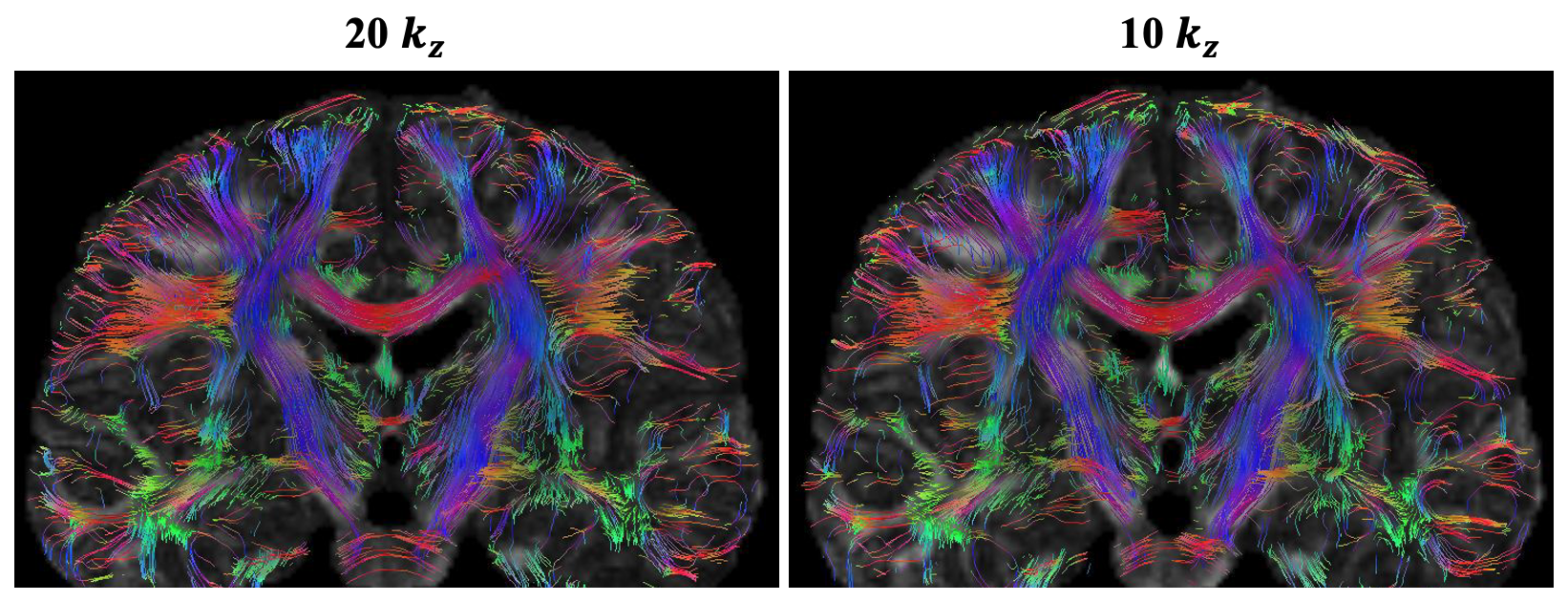}
     \caption{Comparison of fiber tractography from EPEN reconstructions at $N_{kz}$=20 and 10, using same tracking parameters. The slice direction does not show visible deterioration in the quality of the fiber tracts.}
    \label{fig:sequence}
\end{figure*}

\begin{figure*}[t!]
    \centering   
    \includegraphics[width=0.7\linewidth]{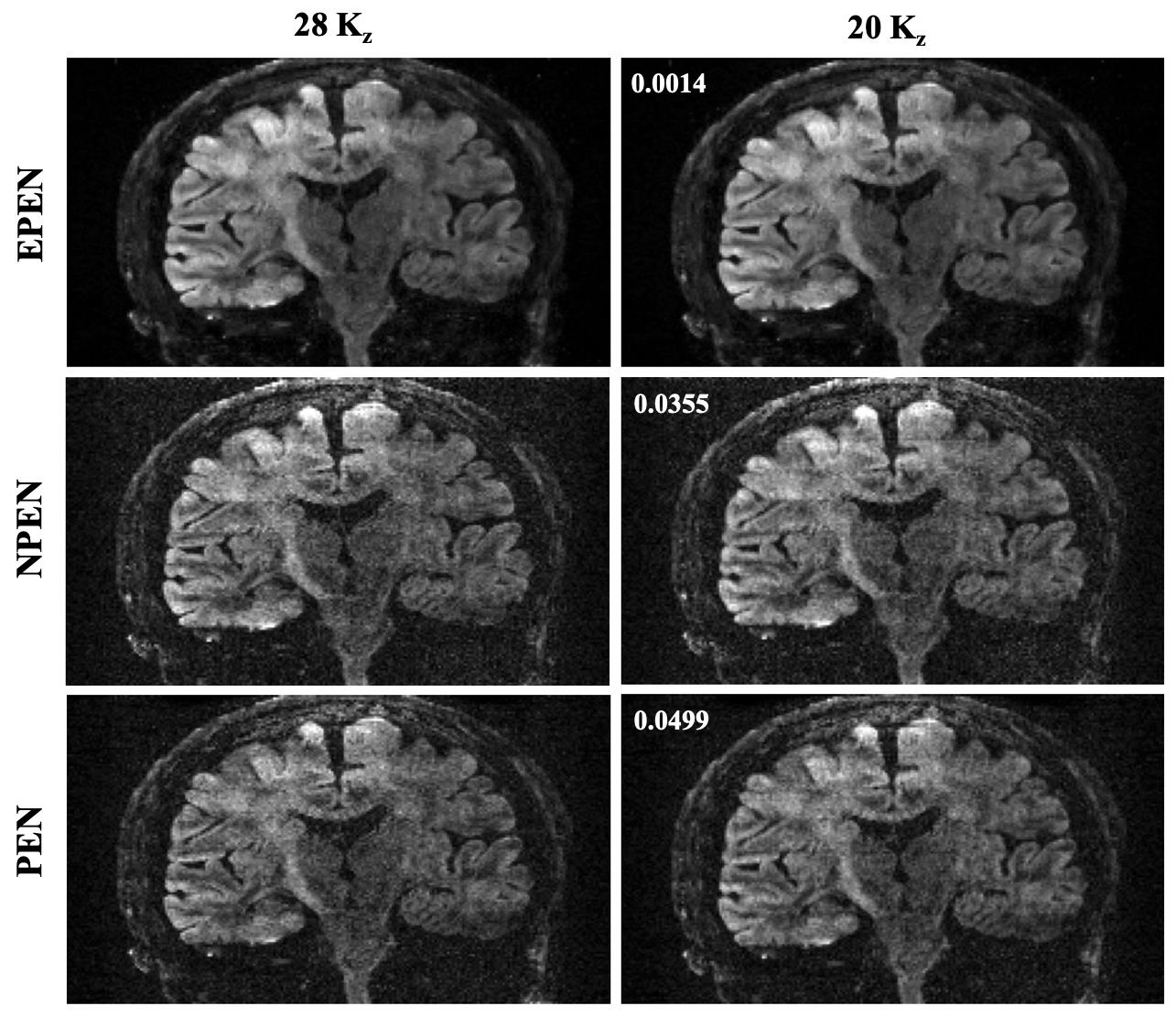}
    \caption{The reconstruction results for the in-vivo dataset 4 with doubled phase encode FOV 28$k_z$, and 20$k_z$ undersampling.}
    \label{fig:sequence}
\end{figure*}

\begin{figure*}[t!]
    \centering   
    \includegraphics[width=0.7\linewidth]{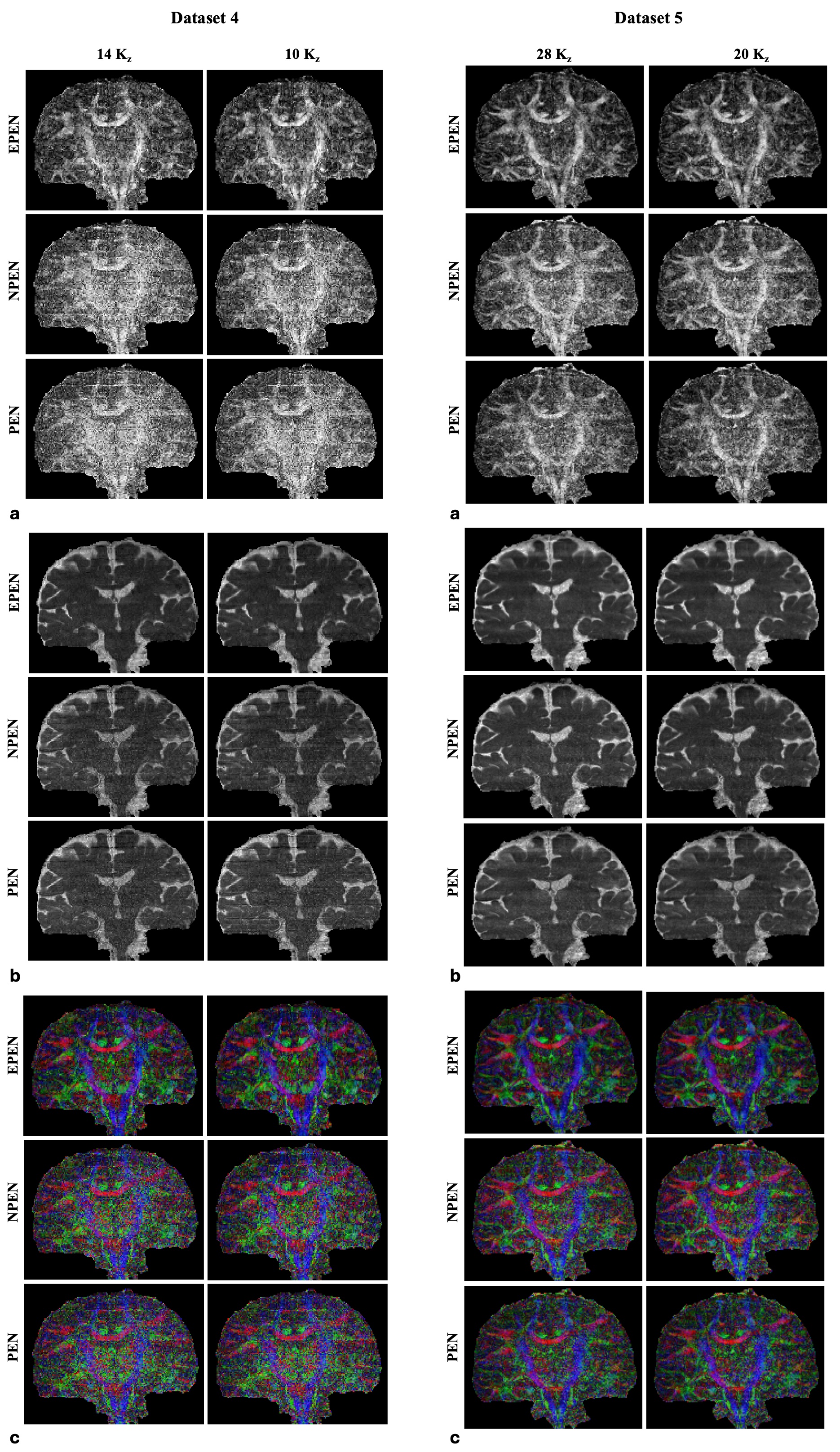}
    \caption{DTI analysis of the reconstruction results for the in-vivo dataset 4 (left) and 5 (right).}
    \label{fig:sequence}
\end{figure*}

\begin{figure*}[t!]
    \centering
    \includegraphics[width=0.7\linewidth]{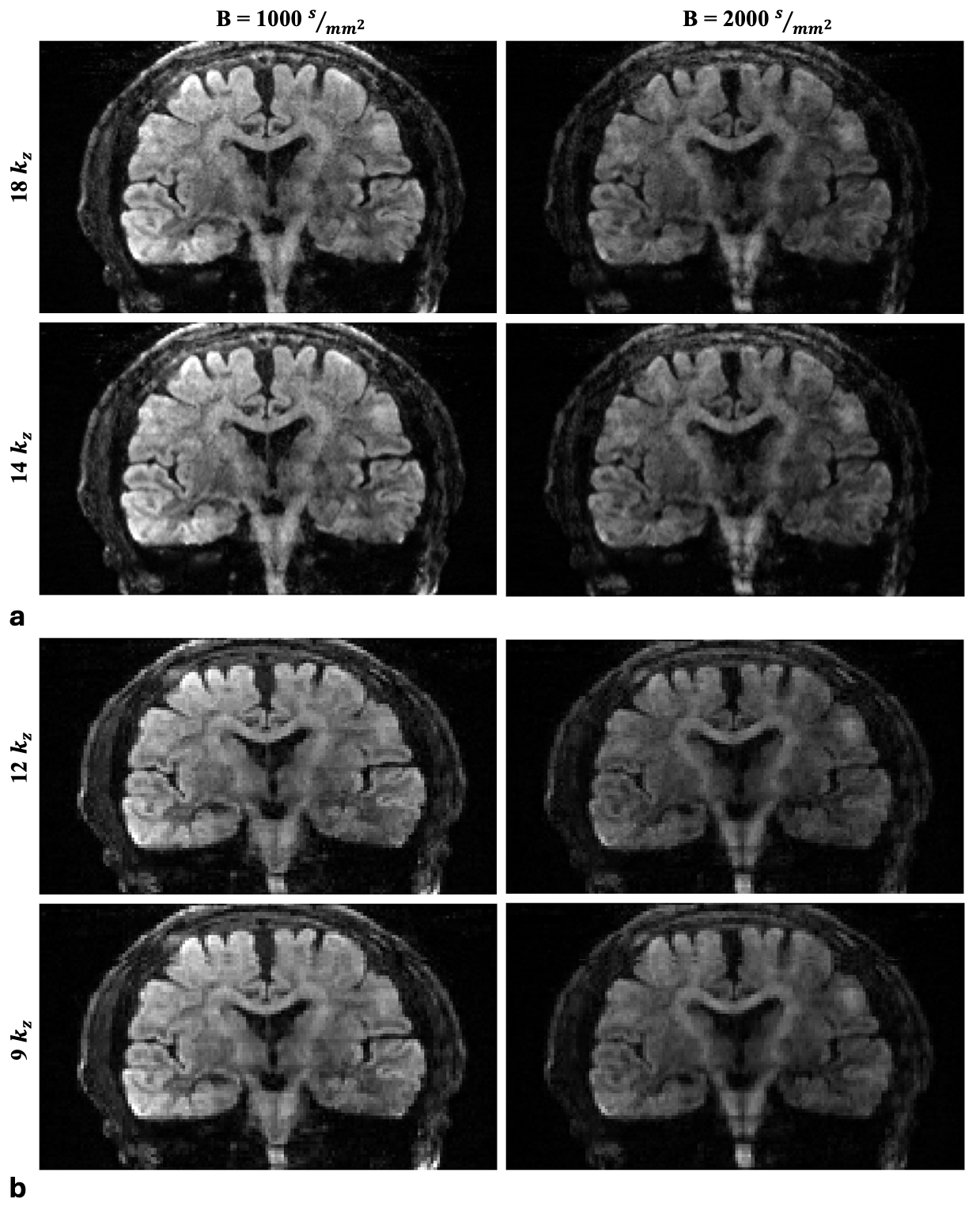}
    \caption{EPEN generalizability across spatial resolutions and diffusion weightings. Reconstructions are shown for datasets 6 and 7 acquired at (a)  1 mm and (b) 1.5 mm isotropic resolution, with b = 1000 $s/mm^2$ and b = 2000 $s/mm^2$. All four reconstructions were obtained using the same energy-model checkpoint trained on the original training described in Section 2.3.1; no retraining or fine-tuning was performed for these new acquisitions. EPEN produces stable, slab-boundary-corrected reconstructions across all four configurations, demonstrating that the learned prior generalizes across substantial changes in spatial resolution and diffusion weighting without requiring additional training data. The RMSE is shown in the upper left-hand corner of the undersampled reconstructions.}
\end{figure*}

\end{document}